\documentclass{osa-article}
\pdfoutput=1
\journal{oe}
\articletype{Research Article}

\usepackage[utf8]{inputenc}
\usepackage{float}

\begin{document}
\title{The point spread function in interferometric scattering microscopy (iSCAT). I. Aberrations in defocusing and axial localization}
\author{Reza Gholami Mahmoodabadi,\authormark{1,2} {Richard W. Taylor},\authormark{1,2} Martin Kaller,\authormark{1,2} Susann Spindler,\authormark{1} Vahid Sandoghdar\authormark{1,2,3*}}
\address{\authormark{1} Max Planck Institute for the Science of Light, Erlangen, Germany\\
\authormark{2} Max-Planck-Zentrum f\"ur Physik und Medizin, Erlangen, Germany\\
\authormark{3} Department of Physics, Friedrich Alexander University, Erlangen, Germany}
\email{\authormark{*} vahid.sandoghdar@mpl.mpg.de}

\begin{abstract}
Interferometric scattering (iSCAT) microscopy is an emerging label-free technique optimized for the sensitive detection of nano-matter. Previous iSCAT studies have approximated the point spread function in iSCAT by a Gaussian intensity distribution. However, recent efforts to track the mobility of nanoparticles in challenging speckle environments and over extended axial ranges has necessitated a quantitative description of the interferometric point spread function (iPSF). We present a robust vectorial diffraction model for the iPSF in tandem with experimental measurements and rigorous FDTD simulations. We examine the iPSF under various imaging scenarios to understand how aberrations due to the experimental configuration encode information about the nanoparticle. We show that the lateral shape of the iPSF can be used to achieve nanometric three-dimensional localization over an extended axial range on the order of 10$\,\mu$m either by means of a fit to an analytical model or calibration-free unsupervised machine learning. Our results have immediate implications for three-dimensional single particle tracking in complex scattering media.
\end{abstract}

\section{Introduction}
Interferometric scattering (iSCAT) microscopy is a powerful label-free technique well suited for the sensitive detection and tracking of nanoscale matter such as individual viruses, colloidal nanoparticles, proteins and single molecules \cite{lindfors2004detection,taylor2019Mini,taylor2019label, young2019interferometric, spindler2016visualization}. While iSCAT is devised for detection of weak scattering inherent to nanoparticles, it belongs to the broader family of interferometric microscopies which include holography, interference reflection microscopy, phase contrast microscopy  and quantitative phase imaging \cite{taylor2019Mini,taylor2019label}. 

In interferometric microscopies, the signal of interest is accompanied by an imaging background which oftentimes is that of a random speckle-like pattern. Such backgrounds present a challenge to remove, especially if they are dynamic. However, with appropriate subtraction or mitigation, ultra-weak signals can be identified down to the shot noise limit. With special sample preparation conditions even single molecules have been detected with iSCAT \cite{celebrano2011single,kukura2010single,piliarik2014direct}. The high detection sensitivity of interferometric microscopy permits localization of single particles to high precision in all four spatio-temporal dimensions, and hence the technique has found application in tracking the fast nanoscale dynamics of various entities \cite{krishnan2010geometry, spindler2016visualization, fringes2016situ, huang2017virus, de2018revealing, taylor2019interferometric}. 

Positional information which is encoded within the interferometric point spread function (iPSF) has been exploited through different methods. In digital holographic microscopy, for example, the positions of large beads are oftentimes extracted through numerical reconstruction of the hologram or through fitting based on Mie theory \cite{Lee07,cheong2010strategies}. In the absence of a model or a fit, experimental point spread functions can be interpreted against a calibration data set, as is common within optical and magnetic tweezing microscopies \cite{gao2017optical, BradacTweezer,de2012recent}. It is only very recently that imaging above and below the focal plane within in-line digital holography has been modeled for an index-matched sample by explicit inclusion of the effect of the lens. Doing so opens the possibility to extend the depth of field to which one can image colloidal particles over a range of many microns \cite{leahy2020large}.

In this work we set out to investigate the iPSF through both theory and experiment. We present a robust vectorial diffraction model to treat the imaging of a nanoparticle arbitrarily positioned about the focus in a wide-field reflection iSCAT microscope. Our model describes the image formation in the presence of aberrations due to the refractive index mismatch between the sample and coverslip, which is common in biological experiments. Furthermore, we discuss the origin of the remarkably long range and fine resolution of iSCAT particle tracking in the axial direction and explain how the iPSF behaves differently from conventional intensity-based PSFs in the diffraction-limited focal region. We also demonstrate how an understanding of the iPSF and application of the model can enable tracking the diffusion of lipids on a large artificial cell under physiological conditions. Finally, we present the use of unsupervised machine learning to extract the axial information encoded in the lateral iPSF profile.

\section{Fundamentals of iSCAT microscopy}
\subsection{The interference equation}
 Central to iSCAT microscopy is the principle that the electric field scattered from the nanoscopic scatterer of interest ($\textbf{E}_{\mathrm{sca}}$) is superposed on the detector with some portion of a reference beam ($\textbf{E}_{\mathrm{ref}}$), rendering the detected intensity as
\begin{equation}
    I_{\mathrm{det}} = {|\textbf{E}_{\mathrm{ref}}|}^{2} + {|\textbf{E}_{\mathrm{sca}}|}^{2} + 2|\textbf{E}_{\mathrm{ref}}||\textbf{E}_{\mathrm{sca}}|\cos{\phi}_{\mathrm{dif}}\,,
    \label{eqn:interf_eqn}
\end{equation}
where ${\phi}_{\mathrm{dif}}\,=\,\phi_{\mathrm{ref}}-\phi_{\mathrm{sca}}$ is the differential phase accumulated between the phases of the scattered and reference fields at the point of detection. The interferometric nature of the cross-term (last term in Eq.\,(\ref{eqn:interf_eqn})) permits detection of extremely weak scattering signals on top of a large background with sensitivity down to the shot noise limit. 

\begin{figure}[htbp]
\centering
\includegraphics[scale=.6]{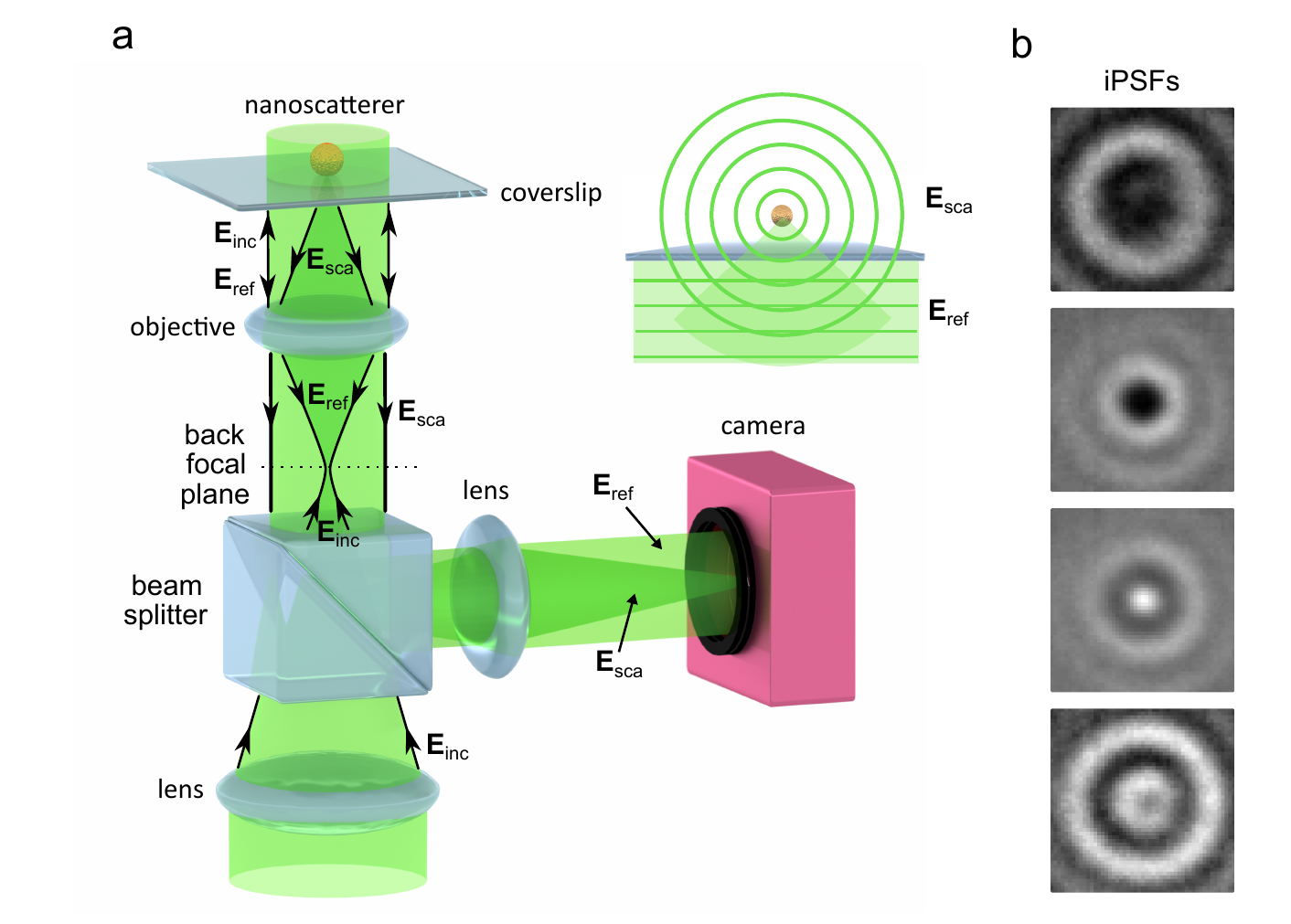}
\caption{(a) Schematic showing the imaging geometry of a wide-field reflection iSCAT microscope. The sample is typically a sub-wavelength scatterer such as a gold nanoparticle, which is taken to be freely positionable with respect to the coverslip and the objective focus. The inset shows that the scattered light with approximately spherical wavefront mixes with planar reflections off the coverslip. (b) Exemplar images on the camera showing the point spread function of a nano-scatterer at different positions. Each image is $2\mu\mathrm{m}\times2\mu\mathrm{m}$ and normalized to the extremum of its own contrast.}
\label{fig:iSCAT_schem}
\end{figure}

The most popular imaging modality for iSCAT is illustrated in Fig.\,\ref{fig:iSCAT_schem}(a), and is composed of a common-path wide-field illumination scheme with camera-based detection. In this mode, the reference beam consists of the reflection of the illumination at the sample interface. The most sensitive imaging is achieved with an oil-immersion objective of high numerical aperture. As it becomes clear below, it is important to realize that these components are typically designed for imaging objects directly placed on the coverslip. 

The iPSFs as shown in Fig.\,\ref{fig:iSCAT_schem}(b) exhibit a characteristic series of rings of increasing size and oscillating contrast caused by the mismatch between the spherical scattered and the near-planar reference wavefronts. These features are important as they encode a wealth of information about the material of the scatterer and its three-dimensional position to nanometer precision. Although we recently benefited from this rich information for tracking transmembrane proteins \cite{taylor2019interferometric}, to date, there has not been a complete description of the iSCAT imaging system. In particular, the important role of omnipresent aberrations on iPSFs has been missing. Such an understanding is critical for accurate and quantitative interpretation of the information contained in iSCAT images.

\subsection{How geometric aberrations contribute to the iPSF and the extended depth of focus}
\label{sec:aberrationa}

The complex iPSF at the plane of detection results from the differing wavefront curvatures of the reflected and scattered fields as well as the phase difference accumulated between them. Typically, the reference field is considered to have a planar wavefront generated by focusing the illuminating light beam in the back focal plane of the microscope objective (see Fig.\,\ref{fig:iSCAT_schem}(a)). The scattered light has a quasi-spherical wavefront emanating from the nanoparticle. Under normal conditions the scattered light is additionally subjected to multiple phase-altering aberrations that arise due to the substrate interface where a refractive index mismatch occurs. The most critical issue in modeling the iPSF is to accurately describe the aberrations introduced by the stratified layers of differing refractive indices which affect both the illuminating and scattered fields. This is not only important for a quantitative description of iSCAT images, but as we shall see, it also provides very valuable axial information. 

\begin{figure}[htbp]
\centering
\includegraphics[scale=.6]{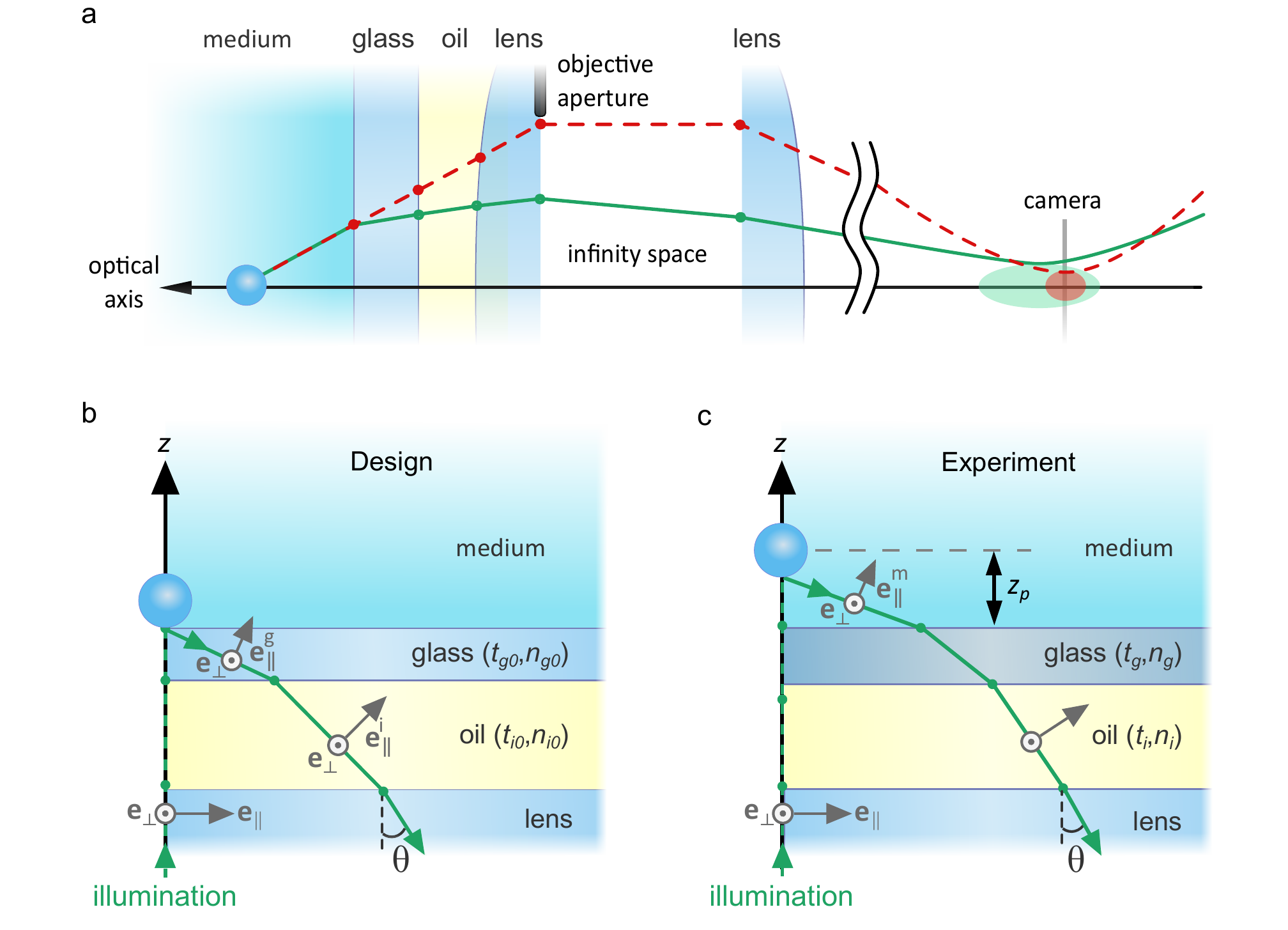}
\caption{(a) Schematic illustrating the role of interface-induced aberrations in perturbing the optical paths of rays which contribute to the point spread function. For a source emitting spherical waves (dashed red ray), the presence of an interface of differing refractive index causes refraction of said rays, shown with a green ray. (b,c) Schematic of the optical paths for the vectorial diffraction model. The intended scenario for nanoparticle imaging wherein the objective achieves an ideal unaberrated performance (b). Here the '0' labels the nominal design values. Note that the perpendicular polarization component of the light is maintained while the rays are refracted by the interfaces opposite to the parallel polarization vector that changes direction for every passing through an interface. In many experimental measurements, the system under study deviates from the intended design scenario, for example by having the nanoparticle situated off the coverslip (c). }
\label{fig:aberrations}
\end{figure}

We start with a qualitative discussion on the origin of the critical aberrations which affect the phase and the amplitude of the scattered field.  Fig.\,\ref{fig:aberrations}(a) illustrates a green ray emanating from a point source such as a nanoparticle scatterer. In the absence of an interface, these rays would seamlessly propagate (dashed red line). On the imaging side, the collected unaberrated rays are focused into a tight diffraction-limited spot which would possess on axis a sharp amplitude profile. In the presence of interfaces, e.g., that of a coverslip, the green ray emitted from the source becomes distorted through refraction (Fig.\,\ref{fig:aberrations}(a)). When tracing such rays on the image side, one finds an aberrated extended focus region because the degree of refraction at interfaces depends on the angle of the wave vectors. 

The traveling phases of the illumination and scattering fields lead to a rapidly varying phase shift, which amounts to a round-trip phase of $\pi$ for a displacement of $\lambda_{\rm m}/4$ where $\lambda_{\rm m}$ is the reduced illumination wavelength in the medium (i.e., in the order of 100\,nm). This phase shift has been previously exploited to achieve an exquisite axial sensitivity in iSCAT \cite{jacobsenbook,krishnan2010geometry,de2018revealing}. In addition, because the wavefront is composed of a distribution of spatially-confined wave vectors, the phase fronts undergo a cumulative $\pi$ shift along the direction of propagation, a phenomenon known as the Gouy phase shift \cite{hwang2007interferometry,feng2001physical}. For the tight focus of the unaberrated beam, the Gouy phase shift occurs rapidly in proportion to the extent of the focal volume. For the aberrated rays, however, the Gouy phase accumulates much more slowly across the focal depth. In the next section, we present a vectorial diffraction model to describe the imaging of a scattering dipole with explicit inclusion of these aberrations for a wide-field reflection interferometric microscope.

\section{A vectorial diffraction model for the iPSF}
The vectorial description of diffraction has proven to be a powerful and accurate framework to analytically compute the light emitted by a dipole through stratified media \cite{torok1997electromagnetic,haeberle2003focusing}. This forms the conceptual basis for our model. We first begin by recasting the interference equation (Eq.\,(\ref{eqn:interf_eqn})) to a vectorial format to accommodate an illumination field with an arbitrary polarization. We write $\mathrm{\textbf{E}}_{\mathrm{inc}} = {{\mathrm{\textbf{E}}_{\mathrm{inc}}}^\parallel}+{{\textbf{E}_{\mathrm{inc}}}^\perp}$, where $\parallel$ and $\perp$ denote two orthogonal polarization states in the plane normal to the propagation direction. The detected intensity is then given by
\begin{equation}
\label{eqn:interf_eqn_vec}
\begin{aligned}
    I_{\mathrm{det}} = &  {|\textbf{E}_{\mathrm{ref}}({\textbf{$\gamma$}})+\textbf{E}_{\mathrm{sca}}({\textbf{$\gamma$}})|}^{2} = {|{{\textbf{E}_{\mathrm{ref}}}^\parallel}+{{\textbf{E}_{\mathrm{sca}}}^\parallel}|}^{2} + {|{{\textbf{E}_{\mathrm{ref}}} ^\perp}+{{\textbf{E}_{\mathrm{sca}}}^\perp}|}^{2} \\
    = &  {|{{\textbf{E}_{\mathrm{ref}}} ^\parallel}|}^{2} + {|{{\textbf{E}_{\mathrm{sca}}} ^\parallel}|}^{2} + 2|{{\textbf{E}_{\mathrm{ref}}} ^\parallel}||{{\textbf{E}_{\mathrm{sca}}}^\parallel}| \cos{{\phi}_{\mathrm{dif}}^\parallel} \\
    & + {|{{\textbf{E}_{\mathrm{ref}}} ^\perp}|}^{2}+ {|{{\textbf{E}_{\mathrm{sca}}} ^\perp}|}^{2} + 2|{{\textbf{E}_{\mathrm{ref}}} ^\perp}| |{{\textbf{E}_{\mathrm{sca}}}^\perp}|\cos{{\phi}_{\mathrm{dif}}^\perp}\,,
\end{aligned}
\end{equation}
where $\textbf{E}$ denotes a vectorial field, and the argument \textbf{$\gamma$} is a vector containing all the physical parameters in the coherent imaging system, which we will shortly discuss and for brevity omit from writing explicitly. 

Experimentally, one often views the iPSF by firstly having subtracted the imaging background (${|{{\textbf{E}_{\mathrm{ref}}} ^\parallel}|}^{2}+{|{{\textbf{E}_{\mathrm{ref}}} ^\perp}|}^{2}$) from $I_{\mathrm{det}}$ and then normalizing again by this background. Hence, we write
\begin{equation}
\label{eqn:iPSFVec}
    \rm{iPSF}  = \frac{{|{{\textbf{E}_{\mathrm{sca}}} ^\parallel}|}^{2} + 2|{{\textbf{E}_{\mathrm{ref}}} ^\parallel}||{{\textbf{E}_{\mathrm{sca}}}^\parallel}|\cos{{\phi}_{\mathrm{dif}}^\parallel} + {|{{\textbf{E}_{\mathrm{sca}}} ^\perp}|}^{2} + 2|{{\textbf{E}_{\mathrm{ref}}} ^\perp}||{{\textbf{E}_{\mathrm{sca}}}^\perp}|\cos{{\phi}_{\mathrm{dif}}^\perp}}{{|{{\textbf{E}_{\mathrm{ref}}} ^\parallel}|}^{2} +{|{{\textbf{E}_{\mathrm{ref}}} ^\perp}|}^{2}}\,.
\end{equation}
In the following, we consider each field contribution in the interference equation.   

\subsection*{The reflected reference field: $\mathrm{\textbf{E}}_{\mathrm{ref}}$}
The reference field arises from a reflection of the incident illumination at the interface between the coverslip and the sample and is, in principle, polarization sensitive. The strength and phase of this field are given by the reflectance \textit{R} and complex amplitude reflection coefficient \textit{$\tilde{r}$} obtained from Fresnel coefficients
\begin{equation}
\label{eqn:Eref}
\begin{aligned}
|{\textbf{E}_{\mathrm{ref}}^{\parallel}}|^2 & = R^{\parallel} |{\textbf{E}_{\mathrm{inc}}^{\parallel}}|^2 \\
{\phi}^{\parallel}_{\mathrm{ref}} & =   \mathrm{arg} (\tilde{r}^\parallel \ {\textbf{E}_{\mathrm{inc}}^{\parallel}})\,.
\end{aligned}
\end{equation}
If multiple interfacial layers exist, one can readily cascade the Fresnel coefficients. 

\subsection*{The scattered field: ${{\textbf{E}}_{\mathrm{sca}}}$}
The amount of light scattered by a nanoparticle is determined by the strength of the induced dipole polarization encapsulated by the polarizability \cite{bohren2008absorption},
\begin{equation}
    {\alpha}_i  = 4 \pi a_1 a_2 a_3 \frac{\epsilon_{1}-\epsilon_m}{3\epsilon_m+3L_i(\epsilon_{1}-\epsilon_m)}\,.
    \nonumber
\end{equation}
Here, the particle semiaxes are \textit{$a_1\leq a_2 \leq a_3$}, $L_i$ denotes the depolarization factor along each axis, and $\epsilon_{1}$ and $\epsilon_{m}$ are the wavelength-dependant permittivities of the nanoparticle and the medium, respectively. For a spherical nanoparticle of radius $a$, $L_i$=1/3, yielding the familiar expression for polarizability,
\begin{equation}
    \alpha  =  4{\pi}{a^3}\left(\frac{\epsilon_{1}-\epsilon_{m}}{\epsilon_{1}+2\epsilon_{m}}\right)
    \label{eqn:polarizability}
\end{equation}
 and the material-dependent scattering phase  
 \begin{equation}
\label{eqn:phaseEs0}
    {\phi}_{\rm{sca}} = \mathrm{arg}(\alpha)\,.
\end{equation}
The scattering cross-section for a spherical particle is then given by 
\begin{equation}
\label{eqn:scatCross}
C\mathrm{_{sca}} = \frac{\textbf{$k$}^4}{6\pi}|\alpha|^2\,,
\end{equation}
where \textbf{$k$} is the wave vector of the illuminating field. It follows that the scattering amplitude for each polarization component is described by
\begin{equation}
\label{eqn:EsAmp}
{E}_{\mathrm{sca}}\,=\,{\eta}T\sqrt{ C_{\mathrm{sca}}}|{\textbf{E}_{\rm{inc}}}|\,,
\end{equation}
where $T$ is the transmittance of the interface. We have introduced \textit{$\eta = 1/\pi\arcsin[\mathrm{min}(\mathrm{NA}/n_{m},1)]$} as a collection efficiency factor to take into account the angular extent to which the scattered light is collected by the objective. To ensure that the collection angle does not exceed the limit of $\pi/2$, the minimum function sets the maximum limit of the $\sin$ function to be 1. Here, we have introduced \textit{$E_{\mathrm{sca}}$} and {$\hat{\mathrm{\textbf{E}}}_{\mathrm{sca}}$} as the complex amplitude and a unit vector for the scattered field $\textbf{E}_{\mathrm{sca}}$, respectively. The former is a complex number accounting for the absolute scattering amplitude and material-specific scattering phase of the nanoparticle. The latter term describes the three-dimensional spatial distribution of the vectorial scattered field. 

We take as our foundation the vectorial diffraction framework for a radiating dipole emitting through numerous stratified layers \cite{aguet2009super}. This treatment accounts for the refractive aberrations introduced in Fig.\,\ref{fig:aberrations}(b) and (c). We begin with a complete description of an unaberrated or `ideal' scenario, depicted in Fig.\,\ref{fig:aberrations}(b) and then apply a correction factor to the phase to account for the effect of aberrations, as formulated by Ref.\,\cite{Gibson1992b}. This correction factor is in the form of an optical path difference (OPD).  It should be emphasized that most modern high-NA objectives are designed to perform aberration-free imaging of a point source seated directly upon the coverslip at the focus of the lens. Hence, we describe this scenario as unaberrated or the nominal design scenario.

Following Ref.\,\cite{aguet2009super}, the vectorial electric field of a radiating dipole situated in the sample medium at the position $\textbf{x}_{p}=(x_p,y_p,z_p)$ with a moment directed along the unit vector $\hat{\textbf{{e}}}_{p}=(\sin{\theta_{p}}\cos{\phi_{p}},\sin{\theta_{p}}\sin{\phi_{p}},\cos{\theta_{p}})$ is given by:
\begin{equation}
\textbf{E}_{m} = E_{m} \ [(\hat{\textbf{e}}_{p} \cdot \hat{\textbf{e}}^{m}_{\parallel}) \   \hat{\textbf{e}}^{m}_{\parallel} + ( \hat{\textbf{e}}_{p} \cdot \hat{\textbf{e}}_{\perp} ) \ \hat{\textbf{e}}_{\perp} ]\,,
\end{equation}
where $\theta_{p}$ and $\phi_{p}$ are the zenith and azimuthal angles of the dipole (the optical axis is taken as z-axis shown in Fig.\,\ref{fig:aberrations}), and $E_{m}$ is the magnitude of the field. The superscript $m$ on $\hat{\textbf{e}}^{m}_{\perp}$ stands for the sample medium. The unit vectors $\hat{\textbf{e}}^{m}_{\parallel}$ and $\hat{\textbf{e}}_{\perp}$ are the \textit{p}- and \textit{s}-polarized components of $ \textbf{E}_{m}$ in the sample medium. We note that since the polarization direction of the \textit{s}-polarized component $\hat{\textbf{e}}_{\perp}$ is preserved in the optical system (see Fig.\,\ref{fig:aberrations}(b)), we do not assign a superscript to it. 

After propagating through the layers in the object space and the objective lens, the dipole field impinging on the aperture is given by
\begin{equation}
\textbf{E}_{a} = E_{m} \ [(\hat{\textbf{e}}_{p} \cdot \hat{\textbf{e}}^{\ m}_{\parallel}) \ {\tilde{t}_{\parallel}}^{\ (1)} {\tilde{t}}^{\ (2)}_{\parallel} \hat{\textbf{e}}^{\ a}_{\parallel} +( \hat{\textbf{e}}_{p} \cdot \hat{\textbf{e}}_{\perp} ) \ {\tilde{t}}^{\ (1)}_{\perp}{\tilde{t}}^{\ (2)}_{\perp} \hat{\textbf{e}}_{\perp}] \ ,
\label{eqn:EfieldAfterLayers}
\end{equation}
where $\tilde{t}^{\ (j)}_{\parallel}$ and $\tilde{t}^{\ (j)}_{\perp}$ are the Fresnel transmission coefficients for \textit{p}-polarized and \textit{s}-polarized light from layer $j$ to layer $j+1$. This takes into account a glass slab for the coverslip and an immersion oil layer with thickness and refractive index pairs $(z_p, n_m)$, $(t_g, n_g)$ and $(t_i, n_i)$ respectively, but it may be generalized to include an arbitrary number of strata. In our case, superscript indices 1 and 2 label the interfaces between water-glass and glass-oil, respectively. The unit vectors of the polarization directions in spherical coordinates are
\begin{equation}
\begin{aligned}
\label{eqn:unitVecSpheCord}
\hat{\textbf{e}}_{\parallel}^{\ m} & = (\cos\theta_{m}\cos\phi,\cos\theta_{m}\sin\phi,-\sin\theta_{m})\\
\hat{\textbf{e}}_{\parallel}^{\ a} & = (\cos\phi,\sin\phi,0)\\
\hat{\textbf{e}}_{\perp} & = (-\sin\phi,\cos\phi,0)\,.
\end{aligned}
\end{equation}
A ray in the immersion layer described by a wave vector $\textbf{k}_{i}$ has the zenith and azimuthal angles of ${\theta_{i}}$ and $\phi$, respectively, yielding
\begin{equation}
\textbf{k}_{i} = -kn_{i}(\cos\phi \sin\theta_{i},\sin\phi\sin\theta_{i},\cos\theta_{i}).
\end{equation} 
The field impinging on the aperture $\textbf{E}_{a}$ (Eq.\,(\ref{eqn:EfieldAfterLayers})) is collected up to the angular extent of NA, and is subsequently imaged on the detector using the Richards-Wolf integral \cite{richards1959electromagnetic,wolf1959electromagnetic,aguet2009super} (expressed in coordinates of the object space),
 \begin{equation}
 \begin{aligned}
\label{eqn:EfieldObjSpace}
\textbf{E}_d & = -iA_{0} \int^{2\pi}_{0} \int^{\alpha_a}_{0} \textbf{E}_{a} \ exp({i\textbf{k}_{i} \cdot \textbf{x}}) \  exp[{ik_{i}\textbf{$\Lambda$}(\theta_{i};\textbf{$\tau$})]\sqrt{\cos\theta_{i}}\sin\theta_{i}} \  d\theta_{i} d\phi \\ & = -iA_{0} \int^{2\pi}_{0} \int^{\alpha_a}_{0} \textbf{E}_{a}  \ exp[{ik_{i}rn_{i} \sin\theta_{i}\cos(\phi-\phi_{d})}]  \\ &  \qquad \qquad \qquad \qquad   exp({-ikn_{i}z \cos\theta_{i}}) \exp[{ik\textbf{$\Lambda$}(\theta_{i};\textbf{$\tau$})] \sqrt{\cos\theta_{i}}\sin\theta_{i}} \   d\theta_{i} d\phi \,.
\end{aligned}
\end{equation} 
Here, \textit{$A_0$} is a scaling factor that incorporates $E_{m}$. The OPD correction term (\textbf{$\mathrm{\Lambda}$}) is given by \cite{aguet2009super}
\begin{equation}
\label{eqn:lambdaOPD}
\begin{aligned}
    \textbf{$\Lambda$}(\theta_i,z_{\mathrm{focus}},z_p,\textbf{$\tau$}) =  & \ n_{i}\cos\theta_{i} \left[ z_{p}-z_{\mathrm{focus}} + n_{i}\left({-z_{p}}/{n_{m}} - {t_{g}}/{n_{g}} + {t_{g0}}/{n_{g0}} + {t_{i0}}/{n_{i0}}\right)\right] \\
    & + z_{p}\sqrt{{n^{2}_{m} - n^{2}_{i}\sin^{2}\theta_{i}}}+ t_{g}\sqrt{{n^{2}_{g} - n^{2}_{i}\sin^{2}\theta_{i}}}  \\
    & - t_{g0}\sqrt{{n^{2}_{g0} - n^{2}_{i}\sin^{2}\theta_{i}}}- t_{i0}\sqrt{{n^{2}_{i0} - n^{2}_{i}\sin^{2}\theta_{i}}}\,,
\end{aligned}
\end{equation} 
\newline
where the vector $\textbf{$\mathrm{\tau}$} \, = \, (n_{i},n_{i0},n_{g},n_{g0},n_{m},t_{i0},t_{g},t_{g0})$ describes the optical parameters of the setup with `0' denoting the nominal design parameters. The field $\mathrm{\textbf{E}}_d$ is a function of the observation point $\mathbf{\mathrm{x}}\,=\, ({-r_d}\cos{\phi_d},{-r_d}\sin{\phi_d},z_{\mathrm{focus}})$ on the detector with ${r_d} = [{{(x-x_p)^2}+{(y-y_p)^2}}]^{1/2}$ and ${{\phi}_d} = \arctan[({y-y_p})/({x-x_p})]$, while $z_{\mathrm{focus}}$ denotes the focal position with respect to the origin of the coordinate system, which lies at the interface between the coverslip and the sample layer. 

To apply this framework laid out for a radiating dipole field to that of a scattering dipole, we make the following modifications. Firstly, we point out that under realistic experimental conditions (i.e., non-design conditions), the incident light traverses an extra optical path given by $[(n_{m}z_{p}+n_{g}t_{g}+n_{i}t_{i})-(n_{g0}t_{g0}+n_{i0}t_{i0})]$ before impinging on the particle. These additional phase terms can be readily included within the OPD to account for the excitation phase. We then define $\textbf{$\Lambda$}_{\rm sca}$ as the aberration term for a  scattering dipole as
\begin{equation}
\label{eqn:lambdaOPDCoh}
    \textbf{$\Lambda$}_{\rm sca} = \textbf{$\Lambda$} + (n_{m}z_{p}+n_{g}t_{g}+n_{i}t_{i})-(n_{g0}t_{g0}+n_{i0}t_{i0}).
\end{equation}
Thus, to obtain the electric field of the scatterer at the detector plane, we consider the electric field $\mathbf{\mathrm{E}}_{d}$ for a radiating dipole (Eq.\,(\ref{eqn:EfieldObjSpace})), replace the OPD term \textbf{$\Lambda$} (Eq.\,(\ref{eqn:lambdaOPD})) with that of a scattering dipole source $\textbf{$\Lambda$}_{\rm sca}$ (Eq.\,(\ref{eqn:lambdaOPDCoh})), and implement an induced dipole moment along the polarization of the incident light $\hat{\textbf{{e}}}_{\rm inc}$ in place of the dipole moment $\hat{\textbf{{e}}}_{p}$\cite{aguet2009super,torok1997electromagnetic,egner1999equivalence}.

\section{Results}
\subsection{The extraordinary extended depth of focus in iSCAT and the uniqueness of the defocused iPSF}
\begin{figure}[htbp]
\centering
\includegraphics[scale=.48]{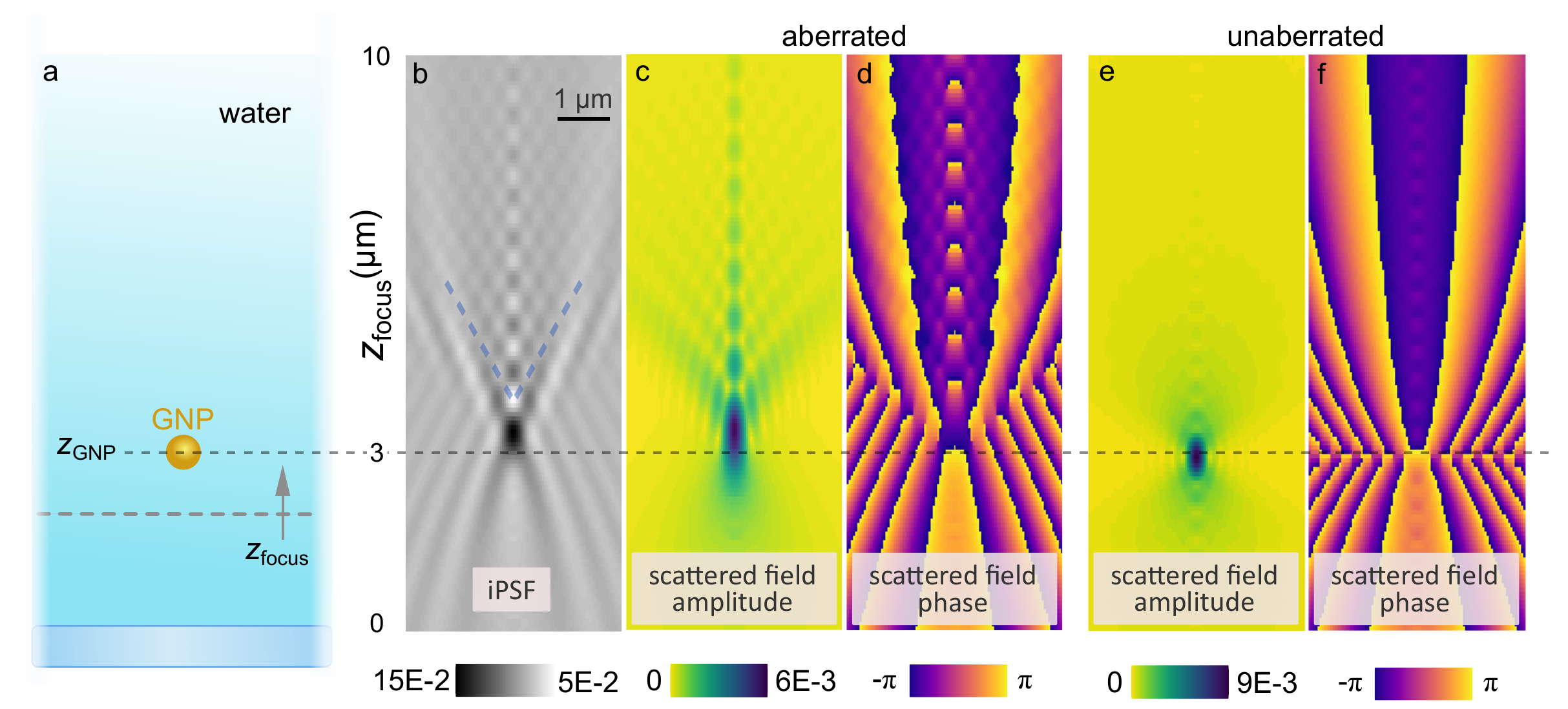}
\caption{The effect of aberrations on the iPSF. (a) The geometry of imaging in which the particle is at a fixed height above the coverslip ($z_{\mathrm{GNP}}\,=\,3\,\mu\mathrm{m}$) and the focal plane is swept starting from the coverslip ($z_{\mathrm{focus}}\,=\,0$) up to \textit{$7\,\mu\mathrm{m}$} above the particle. (b) The meridional profile of the iPSFs as a function of the focal position $z_{\mathrm{focus}}$. With an ever-increasing defocus, additional side lobes begin to appear around the main peak which continue to spread out radially. These features keep the profiles of the iPSFs unique over a long axial range. The blue dashed lines act as a guide to the eye for one such feature. (c,d) The aberrated amplitude and differential phase of the scattered light. (e,f) The corresponding amplitude and differential phase of the scattered light in an ideal interface-free configuration.}
\label{fig:ExtendedFocus}
\end{figure}

Leading on from the qualitative introduction to the role of aberrations introduced earlier in Section \ref{sec:aberrationa}, here we present an explicit examination of the complex scattered light and the interesting features of it such as structure, asymmetry and uniqueness. These aberration-induced contributions richly tailor the iPSF and yield localization of a nanoparticle over an extended axial depth of focus, a feature that is inherent to iSCAT microscopy. We begin by presenting Fig.\,\ref{fig:ExtendedFocus} where we consider the example of a gold nanoparticle (GNP) held above the coverslip and where the focal plane is swept throughout the sample space, depicted in Fig.\,\ref{fig:ExtendedFocus}(a). In Fig.\,\ref{fig:ExtendedFocus}(b-d) we present the meridional iPSF profile, the amplitude and differential phase components respectively for the scattered light for the experimental arrangement shown in Fig.\,\ref{fig:ExtendedFocus}(a). In Fig.\,\ref{fig:ExtendedFocus}(e) and (f) we show the amplitude and the differential phase of the unaberrated scattered field, i.e. the case where the GNP lies in a medium with the refractive index similar to that of the coverslip, hence in absence of any interface.

A notable distinctive feature of the iPSF is the strong asymmetry in the distribution of both amplitude and phase about the position of the GNP. In our formulation, we consider the difference between the phase of the scattered light to that of the reflected light. Therefore, one directly sees the Gouy phase in the phase plots of Fig.\,\ref{fig:ExtendedFocus}. Examining the phase evolution reveals a slow variation when the focus is below the particle position, but this phase then rapidly accelerates close to the true position of the GNP where thereafter one observes a dramatic continuum of ripples in a honeycomb-like pattern that extends onward for many micrometers.  One interesting observation in the amplitude distribution is that the maximum iSCAT signal in the image space occurs for a particle location shortly below the focus position. Moreover, as shown in Fig.\,\ref{fig:ExtendedFocus}(c), the aberrations also affect the absolute strength of the scattered field, which is found to be roughly 40\% lower than its value in the absence of interfaces (see Fig.\,\ref{fig:ExtendedFocus}(e)). The observed asymmetry proves to be a great asset for removing the inherent ambiguity that is caused by the periodic traveling phase along the axial direction in interferometric signals \cite{jacobsenbook,krishnan2010geometry,de2018revealing}. 
  
We note that coverslip-induced aberrations stem from the sample geometry and hence are common not only to iSCAT but also to other microscopy modalities. However, interferometric microscopies uniquely benefit from these aberrations because one detects the scattered signal through its field amplitude rather than intensity, and hence as the former decays more slowly across space, there persists a larger detectable signal over a longer axial range. In the case of intensity-based microscopies with a limited depth of focus (typically of about 0.5$\,\mu$m at high NA), one needs to employ additional phase-based engineering means to extend the axial range \cite{pavani2009three,shechtman2014optimal}.

\begin{figure}[htbp]
\centering
\includegraphics[scale=.75]{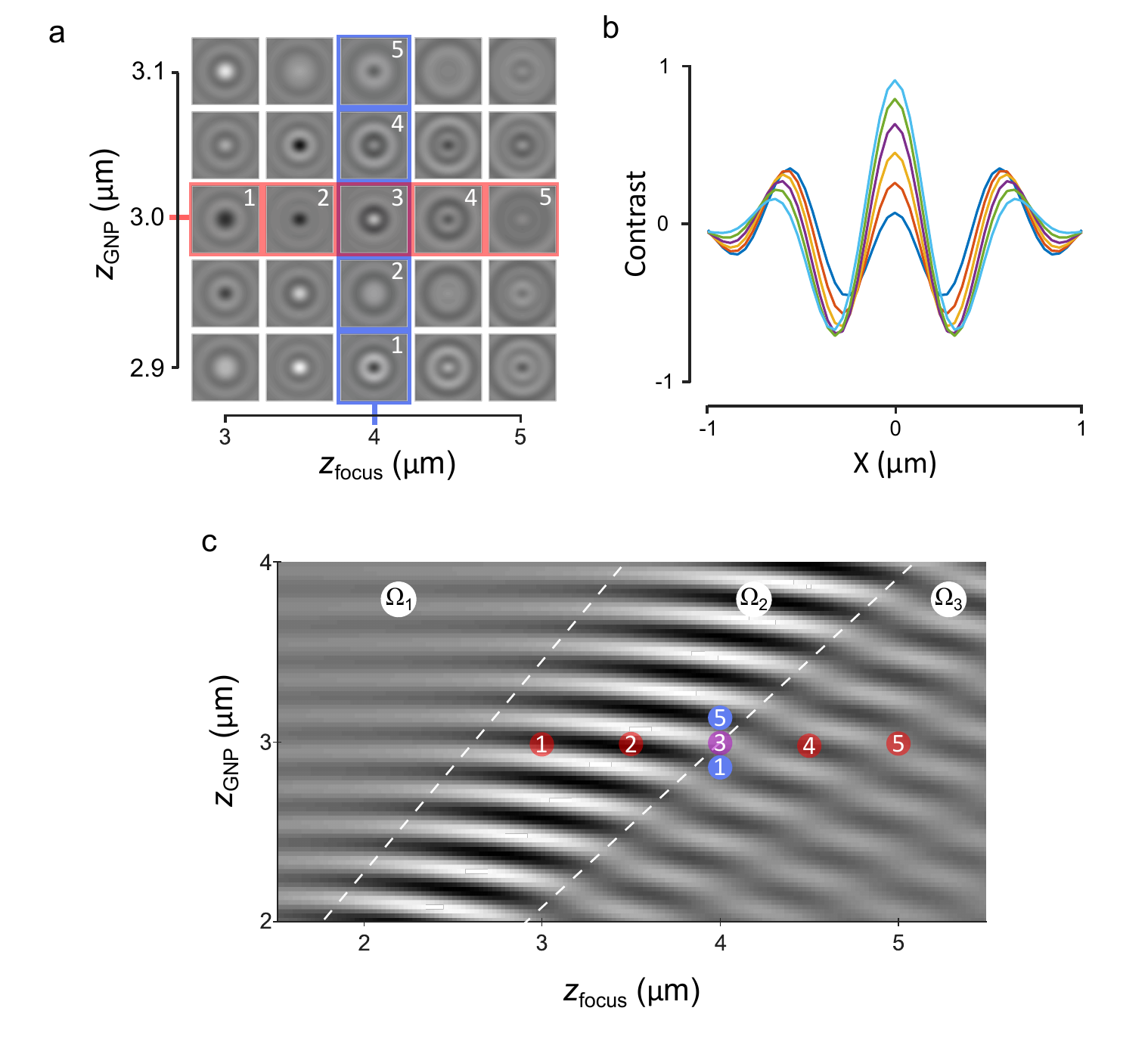}
\caption{iPSFs for different particle and focal positions. (a) Predicted iPSFs from our model as a function of the focal position ($z_{\mathrm{focus}}$) and GNP height above the coverslip ($z_{\mathrm{GNP}}$). The iPSF is an order of magnitude more sensitive to the axial displacement of the GNP compared to changes in the focal plane. Each image is $2\,\mu\mathrm{m}\times2\,\mu\mathrm{m}$. (b) The diametric profiles of several iPSFs are shown for a GNP traveling with the steps of 10\,nm along an axial range of 50\,nm just below $3\,\mu\mathrm{m}$ with the focus being at $4\,\mu\mathrm{m}$. (c) A two-dimensional (2D) map of the contrast of the centroid of the iPSFs as a function of both the height of the GNP and the defocusing of the system. Several iPSFs in (a) are enclosed both in horizontal and vertical directions, in red and blue, respectively and also numbered respectively. The contrast of the centroid of these iPSFs are respectively highlighted in (c). Three regimes of distinct contrast behavior are labeled $\Omega_{1}$, $\Omega_{2}$ and $\Omega_{3}$. Each subfigure is normalized to the contrast extremum of the entire stack.} 
\label{fig:GlamorousFig}
\end{figure}

In particle tracking experiments, the focus of the microscope objective is often set at a fixed height within the sample depth while the particle can travel freely in all three dimensions. We now investigate the iPSF profiles for this configuration theoretically by considering different axial locations of a particle and focus positions, shown in Fig.\,\ref{fig:GlamorousFig}(a). In brief, through defocusing, the corresponding iPSFs accumulate a slowly varying Gouy phase shift in the scattered field over many micrometers, leading to profiles typified by the red-boxed iPSFs in Fig.\,\ref{fig:GlamorousFig}(a). In comparison, when the particle is free to change position the main phase parameter at play is now the accumulated traveling phase between the incident and scattered light with respect to the reference. This phase difference rapidly accumulates over tens of nanometers rendering rapidly changing iPSFs - depicted in blue in Fig.\,\ref{fig:GlamorousFig}(a). In Fig.\,\ref{fig:GlamorousFig}(b), we further highlight the remarkable sensitivity of the iPSF to the axial position of the nanoparticle. Owing to aforementioned aberrations, the radial profiles of the iPSFs show strong distinct structure with axial steps of as small as 10\,nm. This feature was recently exploited to track a gold nanoparticle on the plasma membrane of a live cell \cite{taylor2019interferometric}.

To take a broader view of the parameter space of Fig.\,\ref{fig:GlamorousFig}(a), in Fig.\,\ref{fig:GlamorousFig}(c) we plot the peak centroid contrast as a continuous function of focal position and particle height. In doing so, we can identify three regimes labeled $\Omega_1,\Omega_2$ and $\Omega_3$ that show distinct trends in their contrast dependence. Regime $\Omega_1$ describes the case where the focus is below the particle and displays a weak contrast due to the flat variation in amplitude as well as the accumulation in the Gouy phase being exceptionally slow. Regime $\Omega_2$ displays the strongest contrast as the focus lies close to, and approximately one micrometer above, the physical height of the particle as the aberrated wavefronts begin tightening to a focus which consequently begins to ramp up the accumulated Gouy phase. Finally, there is regime $\Omega_3$ which is similar to the imaging scenario of regime $\Omega_1$ inasmuch as that a large defocus is present, but where now the defocus lies in the space beyond the particle on the non-objective side. As shown in Fig.\,\ref{fig:ExtendedFocus}, this region is richly structured by numerous Gouy phase shift ripples and a slow decay in amplitude that extends outwards for many micrometers far beyond the axial location of the particle. 
To summarize, in interferometric detection scheme, one boosts the useful information encoded by the aberrations in the amplitude and phase of the scattered light by mixing it with a strong planar reference light. In this way, iSCAT yields two advantages. Firstly, the recorded information is doubly imprinted by both the phase and amplitude of the field which is not possible in the intensity-based measurements, and secondly, iSCAT signals win in sensitivity through the structural features of iPSFs. 

\subsection{Verification of iPSF model against experiment and simulation}
We evaluated the calculated iPSF against those obtained from an experimental wide-field reflection iSCAT microscope as depicted in Fig.\,\ref{fig:iSCAT_schem}. This microscope uses an oil-immersion objective (100x, NA\,=\,1.4) and a sample prepared on standard glass coverslip of 170\,$\mu$m thickness. Gold nanoparticles were chosen as model sub-wavelength scattering particles. We used a laser at $\lambda\,=\,$525\,nm for illumination and a camera as detector. Numerical simulations replicating the aforementioned experimental arrangement were performed using finite difference time-domain (FDTD) calculations with the commercially available package 3D Electromagnetic Simulator (Lumerical Inc.).

\subsubsection{A particle moving in the axial direction}
\begin{figure}[htbp]
\centering
\includegraphics[scale=.55]{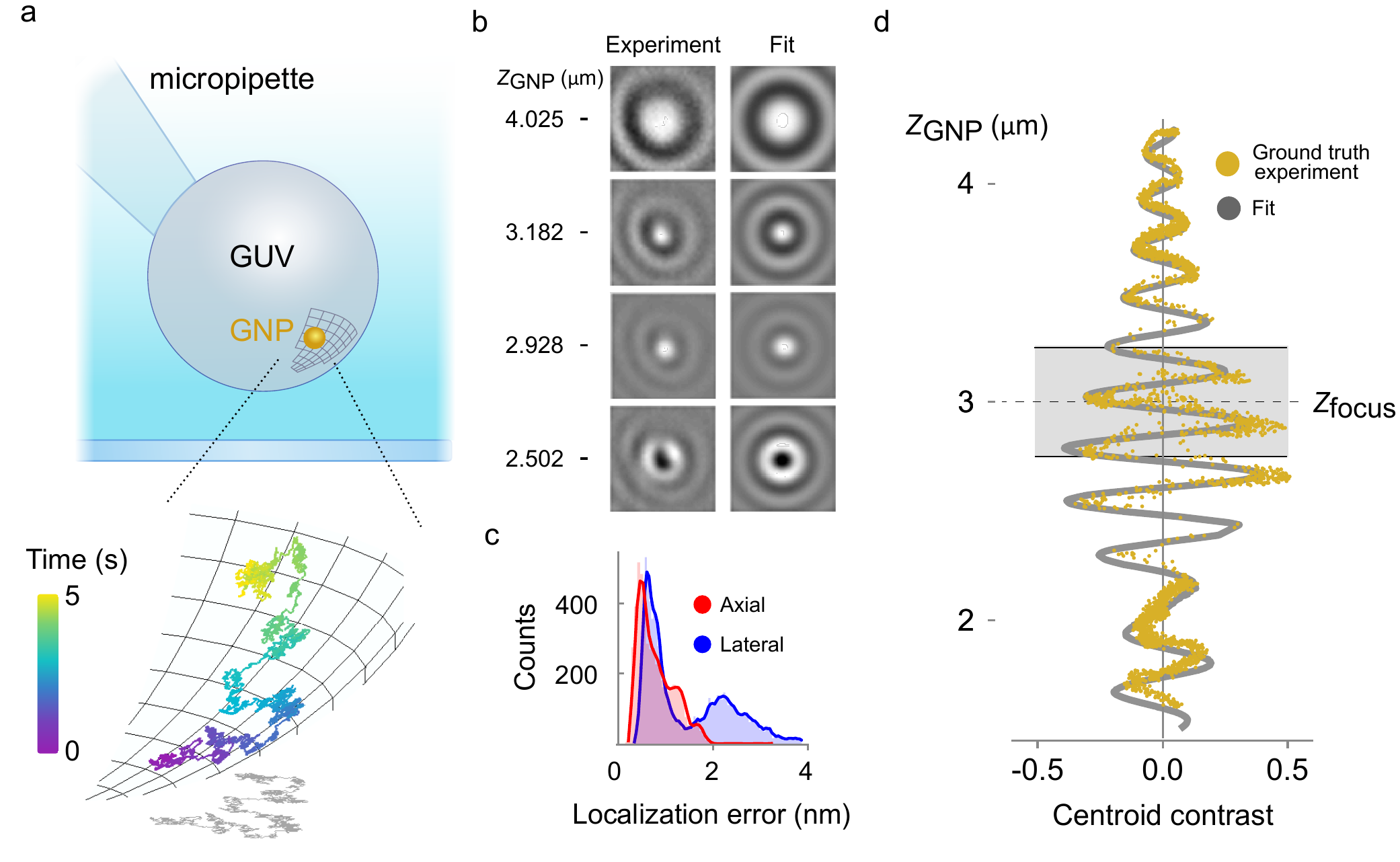}
\caption{Diffusion of a GNP (diameter 40\,nm) upon a GUV. (a) Schematic of the GUV geometry held by a micropipette. Inset: Magnified 3D trajectory of a GNP, with a 2D projection shown in dark gray, as the probe circumnavigates the spherical surface. 
(b) Exemplary iPSFs from the experiment and the corresponding fits to them. Each image is $2\,\mu\mathrm{m}\times2\,\mu\mathrm{m}$ and normalized to the extremum of its own contrast. 
(c) A histogram of the nanometric localization errors in determining the position of the GNP in the lateral (blue) and axial (red) directions by fitting of the model to the experimental iPSFs. (d) The centroid contrast profile of the experimentally measured iPSF of the GNP (mustard) plotted as a function of the `ground-truth' height of the GNP interpreted from the spherical geometry of the GUV. The gray curve shows the result of the fit by the model. The axial range highlighted with a gray transparency denotes the conventional depth of focus achieved in intensity-based imaging.}
\label{fig:GUV_allFitsFig}
\end{figure}

An important application of iSCAT microscopy is in three-dimensional tracking of nanoscopic objects, e.g. in fluidic channels \cite{krishnan2010geometry} or biological cells  \cite{taylor2019interferometric}. Thus, it is important to acquire a quantitative mastery of the iPSF as a function of the axial position of the particle in the sample. As an example of a realistic laboratory application, we tracked a single GNP (diameter 40\,nm) diffusing on a model lipid membrane of a giant unilamellar vesicle (GUV), which is often considered as an artificial cell system \cite{stein2017production}. Here, we investigated a GUV with a diameter of 10\,$\mu$m upon which a GNP was bound and free to diffuse on its spherical surface. As shown in Fig.\,\ref{fig:GUV_allFitsFig}(a), the GUV was held by a micropipette above the coverslip. Further details regarding the procedure to perform single particle tracking on a GUV can be found in Ref.\,\cite{spindler2016visualization, Kaller2020GNP}.

The 3D tracking of the GNP is achieved by fitting the model to the experimental iPSF extracted from a raw iSCAT video. Here, the wide-field image of the GUV delivers direct information on its spherical shape and size, which is used as an independent means to verify the height of the GNP above the coverslip \cite{spindler2016visualization}. The nonlinear iterative fitting of the model can be initiated with a coarse initial guess of the lateral position of the particle. This guess can be obtained, for example, using a  localization algorithm based on radial symmetry \cite{parthasarathy2012rapid}. Similarly, for the initial coarse guess of the axial position of the particle one can either use information from the sample geometry, as is possible here with a GUV, or alternatively one can roughly match the recorded iPSF with that of a model-generated template, such as the series presented in Fig.\,\ref{fig:GlamorousFig}. 

Fig.\,\ref{fig:GUV_allFitsFig}(b) presents several extracted iPSFs and their corresponding fits. We draw attention to the strong agreement between the experiment and model iPSFs. This verifies that indeed the latter encapsulates all of the relevant aberrations which manifest in the experiment to a level sufficiently accurate to describe the sensitive and delicate iPSF formation encountered under realistic experimental conditions. 

Lateral and axial localization of the GNP in each video frame yields 3D trajectories spanning a range of many micrometers in each of the dimensions, as displayed at the bottom of Fig.\,\ref{fig:GUV_allFitsFig}(a). Fig.\,\ref{fig:GUV_allFitsFig}(c) presents a histogram of the axial and lateral localization errors, which amount to only a few nanometers. The bimodal distribution in the lateral localization error (blue, Fig.\,\ref{fig:GUV_allFitsFig}(c)) result from the radial asymmetry in the experimental iPSFs (see Fig.\,\ref{fig:GUV_allFitsFig}(b)). This asymmetry disproportionately hampers the fitting of the model function to the iPSFs which rely upon the secondary lobe information, that is, iPSFs whose central peak is naturally weak. We note, however, that the asymmetry in the experimental iPSFs may originate from other sources of aberration, and if desired could be described within our analytical model description via inclusion of the appropriate Zernike coefficients.

Fig.\,\ref{fig:GUV_allFitsFig}(d) plots the maximum of the central lobe in the iPSF as a function of the axial position of the GNP. As predicted by our model discussed in Fig.\,\ref{fig:ExtendedFocus}, one finds a shift between the axial position of the objective focal plane and the height at which the maximum peak contrast is observed. Furthermore, the focus region has an extended character and the Gouy phase is accumulated in a smooth fashion when the imaging plane is located below the particle. However, the Gouy phase has discrete distortions when the image plane lies above the particle and this in turn leads to the observed irregularities in the oscillation of contrast profile. These data iterate the asymmetry about the imaging plane caused by aberrations and the extended depth of field in iSCAT microscopy. For a visual comparison, we mark the shallower depth of focus achieved with intensity-based imaging from high NA objectives with the gray band in Fig.\,\ref{fig:GUV_allFitsFig}(d). 

We note that in determining the ground truth for height assignment in the experimental trajectory, we assume the GUV to be a perfect sphere. The GUV, however, does not possess a uniform radius of curvature. This leads to slight disagreement with the height assignment found via the central contrast of the modeled iPSFs for GNP positions in larger defocusing regions.

\begin{figure}[htbp]
\centering
\includegraphics[scale=.55]{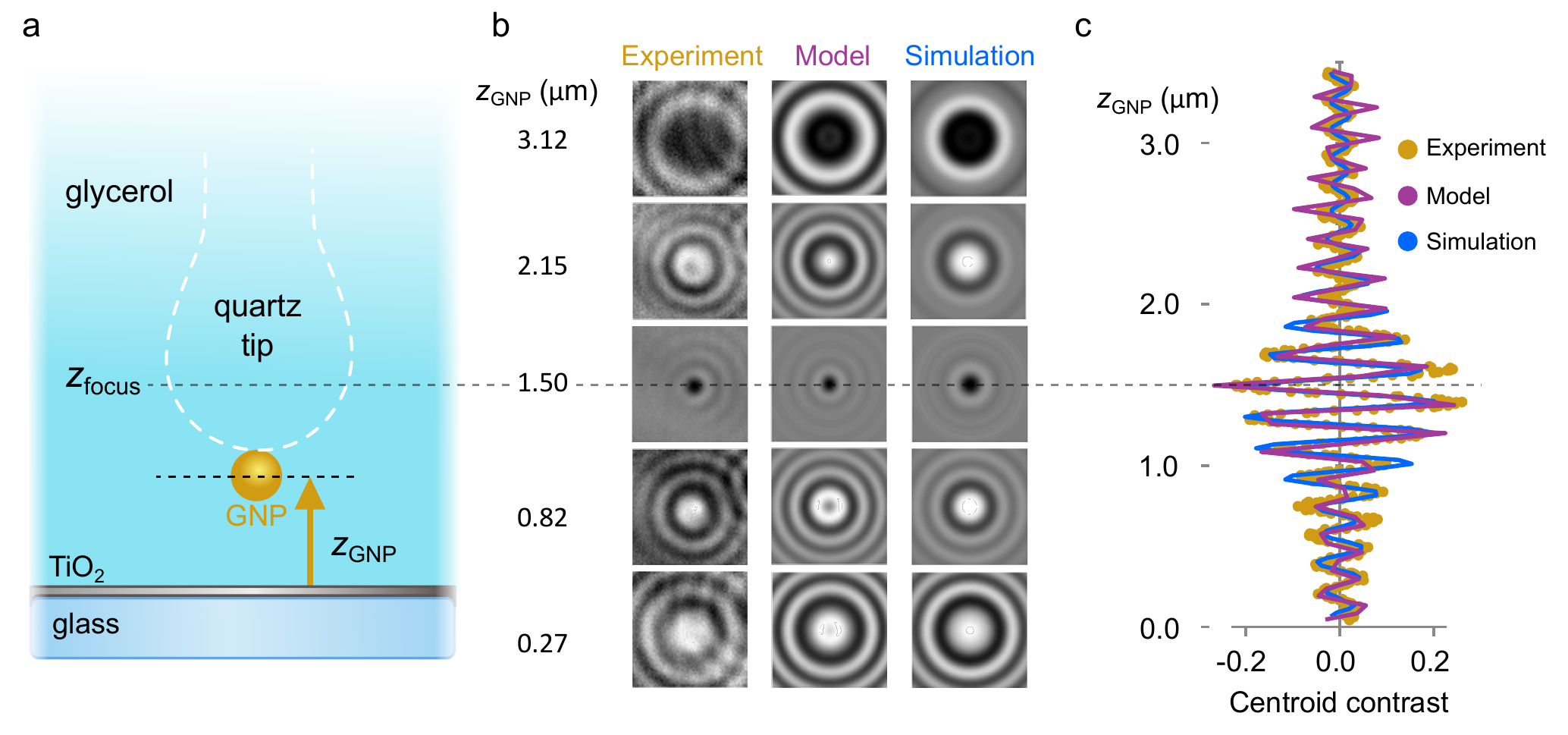}
\caption{The iPSF of a GNP (diameter 50\,nm) that is positionable axially against a static focal plane ($z_{\mathrm{focus}}\,=$1.5$\,\mu$m). (a) Schematic showing a GNP mounted upon a rounded quartz tip immersed in index-matching glycerol. (b) iPSFs from selected heights showing mutual agreement between experiment, model and full FDTD simulation. Each image is $2\,\mu\mathrm{m}\times2\,\mu\mathrm{m}$ and normalized to the extremum of its own contrast. (c) Comparison of the central contrast of the iPSF from the experiment (mustard) described in (a) and with those predicted with our model (violet) and simulation (blue).}
\label{fig:tip_TiO2_retrievingFig}
\end{figure}

To avoid the complications of realistic systems, we also devised a textbook scenario, where a nanoparticle could be positioned only in the axial direction under full control. Here, we placed a GNP (diameter 50\,nm) upon the 50\,$\mu$m tip of a rounded quartz tip fabricated by melting a quartz rod (diameter 1\,mm) using an oxyacetylene flame to produce a rounded end (see Fig.\,\ref{fig:tip_TiO2_retrievingFig}(a)). A dilute solution of GNPs was drop cast upon the plasma-cleaned quartz tip which resulted in several GNPs being deposited across the spherical tip surface at a low density. The quartz tip was independently positionable with respect to the coverslip by a piezoelectric actuator while both the coverslip and objective focus remained static throughout (see Fig.\,\ref{fig:tip_TiO2_retrievingFig}(a)). To suppress unwanted scattering from the quartz tip, the sample medium was index matched to quartz ($n_{\mathrm{med}}$\,=\,1.461) by use of a 91.5\% (w/v) glycerol solution. We remark that a faint background signal from incomplete suppression of the tip scattering remains, but it does not affect extraction of the iPSF. A thin layer (25\,nm) of TiO$_{2}$ was deposited upon the coverslip via atomic deposition to compensate the reduction in reflectivity from the coverslip interface. We note that the added $\mathrm{TiO_{2}}$ layer is thin enough to not perturb the wavefronts. The focus of the microscope objective here was positioned at a height 1.5$\,\mu$m above the coverslip. 

A stack of measured, modeled and simulated iPSFs is shown in Fig.\,\ref{fig:tip_TiO2_retrievingFig}(b) for GNPs placed at various axial positions. To examine the agreement among these three case studies, we plot the central contrast of each iPSF over the full axial range in Fig.\,\ref{fig:tip_TiO2_retrievingFig}(c). Again, the degree to which all three sets agree with one another reflects the accuracy of the model despite the high sensitivity of the aberrations to the details of the experimental arrangement such as thicknesses and indices of the individual layers. We emphasize that even for the iPSFs with a weak peak contrast, which may often be regarded as giving no detectable signal, the side lobes are highly structured and provide a readily usable alternative. We also draw attention to the comparison of the axial contrast trend here to that of Fig.\,\ref{fig:ExtendedFocus}, with a greater symmetry in the distribution of contrast above and below focal plane ($z_{\mathrm {focus}}\,=\,1.5\,\mu$m). This is due to the higher refractive index of glycerol, which is more closely matched to that of the glass coverslip than that of the water, hence weakening the effect of aberrations. This effect is also evident in the better match between the objective focal plane and the height at which the maximum iPSF contrast takes place. We point out, however, that an extended focus is still maintained. 

It is also meaningful to compare the model against the simulation. While both agree compellingly with the experimental results, we note numerous advantages to a model framework over simulation. The model offers a physical picture of the formation of the iPSFs, including all the layers of varying refractive indices in a real experimental configuration and regardless the thickness of these interfacial layers, it samples the wave vectors at the exit pupil and analytically calculates the associated aberrations. The computational time for the model is negligible (milliseconds) whereas a full electromagnetic simulation requires times on the order of hours to days. Simulations are particularly helpful if the geometry of the system is complex \cite{Lin2020}. However, in the simulation of the electromagnetic fields over a range of several micrometers, one must sample the field close to the object and then numerically propagate it further owing to the expensive computational resources and time this would otherwise require. Indeed, simulations might be susceptible to sampling artifacts and convergence problems for challenging sample geometries. In our case, the limited accuracy of our simulations can be seen in a somewhat poorer agreement with the experiment than when compared to the results of the model (see Fig.\,\ref{fig:tip_TiO2_retrievingFig}(c)).

\subsubsection{Focusing through a nanoparticle held at a fixed height}
\begin{figure}[htbp]
\centering
\includegraphics[scale=.55]{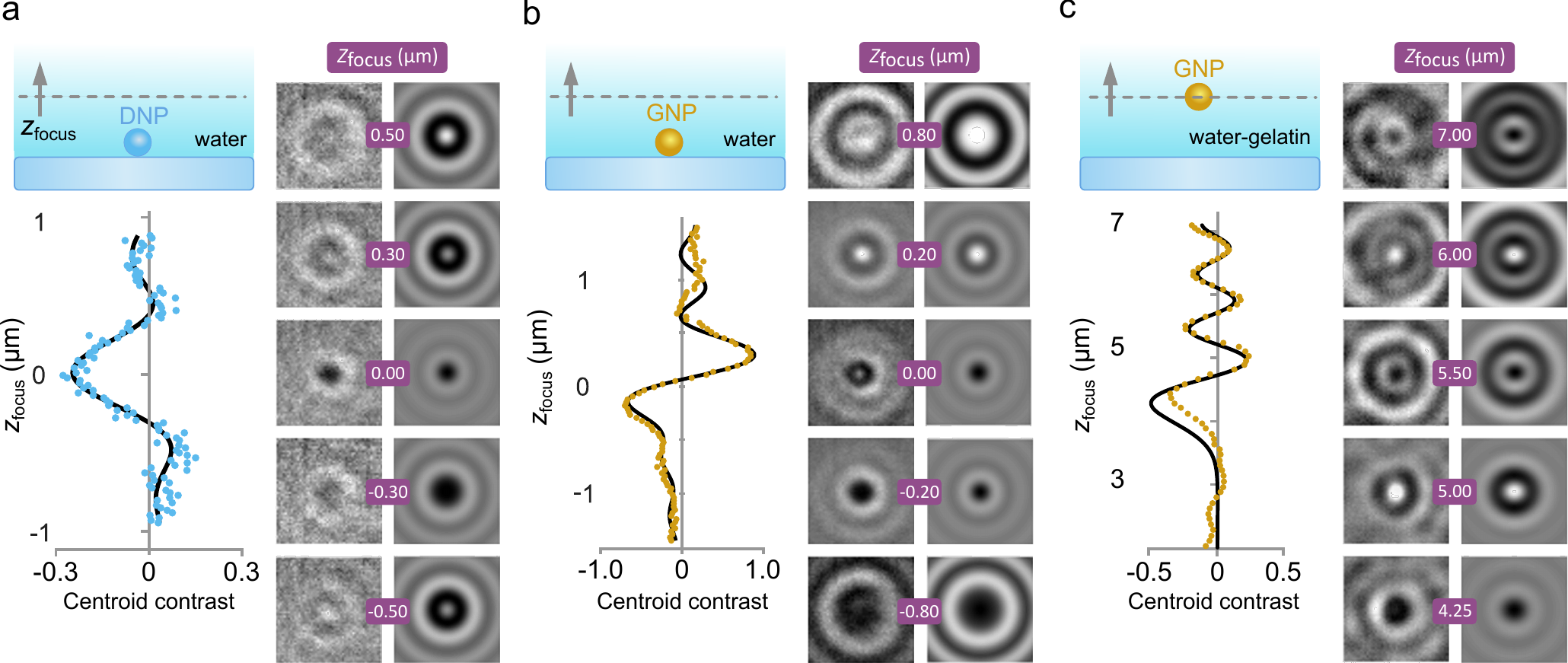}
\caption{Contrast variation of the central lobe of the iPSF for a nanoparticle situated upon the coverslip and swept axially through the objective focus. (a) A dielectric nanoparticle in water. (b) A gold nanoparticle in water. A representative set of iPSFs at selected heights are also presented for the experiment (left column) and as predicted by the model (right column) with design parameters $(n_{i0},n_{g0},t_{i0},t_{g0})=(1.5,1.5,100\,\mu\mathrm{m},170\,\mu\mathrm{m})$ and the fit results $(n_{i},n_{g},t_{g})=(1.5018,1.499,184\,\mu\mathrm{m})$ for (a) and $(n_{i},n_{g},t_{g})=(1.5023,1.4985,179\,\mu\mathrm{m})$ for (b). (c) The same scenario as in (b), but for a static GNP situated off the coverslip. The height of the particle was fitted to be $z_{\mathrm{GNP}}\,=$4$\,\mu$m, in water. The medium here was a dilute mixture of gelatin in water with $n_{m}=1.33$. Each image is $2\,\mu\mathrm{m}\times2\,\mu\mathrm{m}$ and normalized to the extremum of its own contrast.}
\label{fig:defocus}
\end{figure}

Another situation of practical interest concerns the case where the focus of the microscope objective is tuned through the sample. To study this situation in a model system, we examined a GNP placed at the water-glass interface as well as a GNP embedded within gelatin at 4\,$\mu$m above the coverslip. Furthermore, to investigate the effect of the material-dependent phase shift upon scattering, we also compared these results with those of a dielectric nanoparticle at the water-glass interface (see Fig.\,\ref{fig:defocus}). 

Panel (a) of Fig.\,\ref{fig:defocus} considers the elemental case of a dielectric particle upon the coverslip, e.g. as in applications of protein detection \cite{piliarik2014direct}. Here, it is important to understand the iPSF in order to adjust the focal plane during measurements or to maximize the detected contrast. We find that when the focal plane coincides with the position of the dielectric particle sitting on the coverslip, the contrast is maximally destructive, testifying to the Gouy phase shift, wherein the focused point source of light is out of phase with a plane wave of initially the same starting phase by $\pi/2$. Here, it should be borne in mind that since the imaginary part of the dielectric function for the particle is negligible, the phase of the scattered field is the same as that of the illumination. We also observe, accordingly, that the observed contrast has a strong symmetry about the position of the particle. 

Panel (b) considers a GNP, which introduces an additional material and wavelength-dependent phase shift, $\phi_{\mathrm{sca}}$ on the incident light. This shift renders a GNP with a more weakly negative contrast in the focus as compared to the dielectric particle in (a), requiring defocusing to maximize the contrast. In panel (c) we consider the case of a GNP displaced off the coverslip, thus invoking the need for the introduction of additional phase-shifting aberrations. Indeed, as observed in Fig.\,\ref{fig:ExtendedFocus}, we find that the envelope of the amplitude is asymmetric, and the contrast experiences many cycles over the axial range, while decaying much more slowly with progressive defocusing above the position of the particle. 

Upon closer examination, the axial trends of the iPSFs presented in panels (a) and (b) of Fig.\,\ref{fig:defocus} also reveal some asymmetry about the focus position. These are due to slight deviations of the coverslip and immersion oil properties (thickness, refractive index) with respect to the design parameters (see details in the caption of Fig.\,\ref{fig:defocus}). The asymmetry observed in panel (c), which is due to the aberrations caused by defocusing through the water-glass interface, however, is much more pronounced. We wish to emphasize that in a truly aberration-free microscope, the Gouy phase shift and the amplitude distribution are symmetric about the focus and hence lead to ambiguity in height assignment \cite{deschout2014precisely,zhou2019advances}.

\subsection{Feature-based height-assignment of iPSFs through machine learning}
Conventional methods such as iterative algorithms widely used in interferometric imaging techniques, e.g. in digital holography, achieve accurate results at high processing speeds for strong signals from large particles (typically micron-sized). However, they tend to face difficulties tackling noisy distorted signals and have, thus, employed machine learning tools for various image analysis tasks \cite{hannel2018machine,shao2020machine}. Considering that efficient and accurate 3D tracking of nano-matter has enormous potential in modern biology, biophysics and soft condensed-matter physics, we now discuss methods to characterize measured iPSFs from small nanoparticles. The iPSF vectorial model developed in the previous section can now be used as a fit function for experimental measurements using a nonlinear optimizer. This would allow one to deduce the position of the nanoparticle in all three spatial dimensions. However in the presence of a background and a high level of noise in the iPSF signal, axial tracking becomes challenging. Machine learning techniques perform reliably in such conditions. Moreover, as opposed to nonlinear fitting algorithms, such learning-based tools do not require initial guesses of the axial position of the particle.

\begin{figure}[htbp]
\centering
\includegraphics[scale=.55]{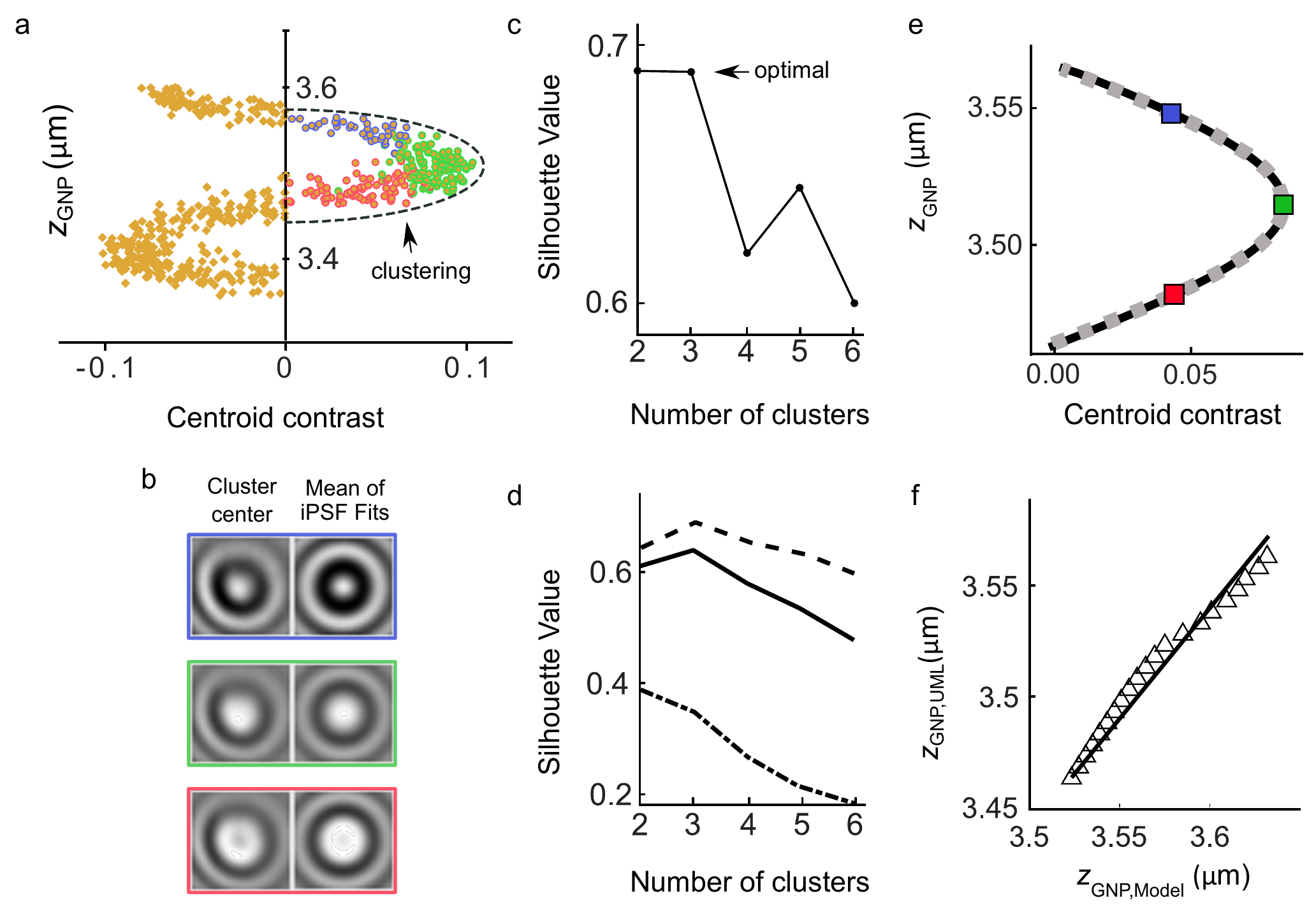}
\caption{Machine learning-assisted height assignment of iPSFs. (a) The centroid contrasts of the iPSFs for a GNP on a GUV are plotted against the verified height above the coverslip. The periodic cosine distribution of contrasts makes the height assignment ambiguous if the centroid contrast is solely used. A K-means clustering algorithm applied to the iPSF profiles from within the black-dash region finds three distinct classes of iPSFs, labeled in blue, green and red. Inspection reveals this classification sorts the iPSFs into self-consistent height-specific classes. (b) The mean iPSF shape within each class, showing class-distinct features and hence confirming the validity of classification presented in (a). Each image is $2\,\mu\mathrm{m}\times2\,\mu\mathrm{m}$ and normalized to the extremum of its own contrast. (c) The silhouette coefficient as a function of the number of chosen clusters. The optimal number of cluster classes in this example is found to be three. (d) The silhouette value curves of synthesized iPSFs with the model for the same axial range with three different SNRs of 1,\,2 and 4 shown with dashed-dotted line, solid line and dashed line, respectively. A simple cosine function is used to assign the relative height to the clustered iPSFs, which here is plotted with a dashed gray line in (e). The cosine function has a periodicity of $\lambda_m/2$ and passes through the iPSF centroid value of the green cluster at its peak. The 2D shape of the iPSFs for all other heights can be estimated using the clustered iPSFs. The centroid contrasts of the estimated iPSFs are plotted with the solid black line. These estimated iPSFs can be then fitted using the model to obtain their axial position which are plotted against each other in (f) with triangular markers. A 45 degree line overlaid on the markers shows a nanometrically precise linear trend.}
\label{fig:clusteringFig}
\end{figure}

We recall that the challenge in tracking particles axially in earlier iSCAT studies was the short accessible axial range of about 100\,nm limited by the periodicity of the contrast in the central lobe of the iPSF: when attempting to assign a height to the centroid contrasts (see the region outlined with the dashed line in Fig.\,\ref{fig:clusteringFig}(a)), one finds multiple solutions for the axial position of the GNP, $z_{\mathrm{GNP}}$. In Fig.\,\ref{fig:GlamorousFig}(b), however, we showed that the ring features of the iPSF provide an exquisite sensitivity to axial displacements. We now show that this sensitivity can be exploited in an unsupervised machine learning scheme to resolve the motion of a GNP. In Fig.\,\ref{fig:clusteringFig}(a), we present an example of a GNP that travels axially a known distance of 300\,nm along the optical axis. In Fig.\,\ref{fig:clusteringFig}(b), the iPSFs associated with the three distinct color-coded regions of Fig.\,\ref{fig:clusteringFig}(a) illustrate how their unique ring structure resolves this ambiguity. We now investigate the application of unsupervised machine learning tools for nanoparticle tracking. The proof of principle work shown here can be extended to even longer axial ranges using more complex deep learning algorithms. We therefore quantify the natural branching in the iPSFs to facilitate benchmarking for the development of follow-up learning algorithms. 

The first step of the procedure is to group similar looking iPSFs. This can be done by defining a so-called distance function as a quantitative measure to compare different iPSFs. This can be done by computing the pixel-wise difference between two iPSFs under study and then summing up the squares of the outcome for all pixels. We then employ a K-means-based unsupervised clustering algorithm \cite{lloyd1982least} to sort all iPSFs within the region of interest (the region outlined with the dashed line in Fig.\,\ref{fig:clusteringFig}(a)) by mutual similarity into an optimal number of groups - referred to as `clusters'. In Fig.\,\ref{fig:clusteringFig}(a), we color-code three such clusters. Here we used the default K-means++ algorithm \cite{david2007vassilvitskii} from the Statistics and Machine Learning Toolbox in MATLAB. For each cluster, the iPSFs are averaged to obtain their `cluster center'. The group of iPSFs are then fitted with the model and averaged within each cluster to obtain their mean fit iPSF. In Fig.\,\ref{fig:clusteringFig}(b), we plot the three cluster centers next to the mean fitted iPSFs for comparison. The clustered iPSF representatives show excellent agreement with the outcome of the complete physical model. We note that because the iPSFs are radially symmetric, we can extract the mean radial profile of the iPSFs over $N\times N$-pixels to reduce the complexity of this computation from $N^2$ down to $N/2$ and, thus, significantly boost the computational performance with a minimum memory footprint. After a successful clustering we can go from 1D profiles back to the full 2D iPSF shapes while using the labels and groupings that we obtained for the 1D case. Interestingly, the results presented in Fig.\,\ref{fig:clusteringFig}(b) show that we can nevertheless go beyond the structural rings of the iPSF to even account for slight imperfections in the imaging setup.

Next, we benchmark the clustering accuracy through a `silhouette value' as a measure for the consistency within clusters \cite{rousseeuw1987silhouettes}. The silhouette value of one data point after being clustered is defined as the difference between its smallest mean inter-cluster dissimilarity and its mean intra-cluster dissimilarity. The former quantity is a measure of how dissimilar (i.e., far) this point is compared to the other points in other clusters than its own cluster. The latter quantity denotes the average dissimilarity of this point compared to all other data points within the same cluster. The silhouette coefficient is normalized to $\pm\,1$. When an iPSF is strongly matched to its own assigned cluster, but poorly matched to other clusters, the silhouette coefficient tends to $+1$. If a silhouette value is larger than 0.5, one considers the data to tend to cluster. In Fig.\,\ref{fig:clusteringFig}(c), we present the silhouette coefficients obtained when sorting the experimental data presented in panel (a) with an increasing number of cluster groups. The curve starts at a value about 0.7 and then shows a sudden drop when more than three cluster groups are considered. This suggests the optimal number of iPSF clusters to be three. In this case, the SNR of the experimental iPSFs were about 2. 

To examine the robustness of this approach to the SNR, we synthesized iPSFs for the same axial range and added different noise levels. We performed three rounds of modeling with SNR levels of 1,2 and 4 and followed it with the silhouette analysis shown in Fig.\,\ref{fig:clusteringFig}(d). The dashed-dotted curve shows the SNR case of 1. In this case, the silhouette value starts with a value below 0.5 and falls almost linearly suggesting that there is no meaningful grouping in the iPSFs as all the discriminatory features between them are buried in noise. The solid line presents the case for a SNR of 2. The curve starts with a silhouette value of about 0.6 and descends for more than three clusters, similar to what we observed for the experimental data in panel (c). The case of a higher SNR of 4, yields the silhouette curve shown with the dashed line. The curve suggests that the discriminatory features of iPSFs at this SNR are strong enough to show a meaningful grouping even for more than three clusters. We note that for higher SNR levels more iPSF side lobes lie above the noise level and provide additional discriminatory sources of information. 

The iPSFs extracted through clustering can be used together with a simple cosine function to build a template for the iPSFs in a given axial range. The iPSF centroid value of the green cluster is solely used as the maximum value of the cosine term $\cos(2 k_{m} z_{\mathrm{GNP}})$ to obtain the dashed gray line shown in Fig.\,\ref{fig:clusteringFig}(e). The blue and red iPSFs sit opposite to each other on the guiding curve due to their dissimilarity in the clustering process. Using the 2D shapes of the clustered iPSFs and their relative axial distance, one can interpolate and extrapolate iPSFs for every single nanometer of this axial range. The centroid contrasts of the estimated iPSFs plotted with the solid black line overlaps strongly with the guiding dashed gray line. The 2D estimated iPSFs can be fitted independently using the iPSF model to obtain the axial position of the particle. In Fig.\,\ref{fig:clusteringFig}(f), we plot the results from the unsupervised machine learning method against the height of the particle using the model, revealing an excellent agreement. 

\section{Discussion and outlook}

Optical microscopy has played a very important role in the development of many disciplines of nanoscience and nanotechnology over the past three decades and continues to be a dynamic field of research \cite{weisenburger2015light}. One of the most valuable modern methods in optical microscopy is the realization that one can localize single molecules and nanoparticles with much better precision than the diffraction limit, thus, giving access to the investigation of nanoscopic trajectories. While most of the efforts in single particle tracking have been limited to lateral movements on surfaces, there is an acute need for three-dimensional trajectories, as methods become powerful enough to study more complex geometries such as live biological cells and tissues. The intrinsically confined nature of the axial focus profile of the fluorescence intensity and the low SNR in single-molecule fluorescence microscopy render such endeavors very challenging. One of the approaches to combat this problem is based on PSF engineering, where one sculpts the PSF to provide axial information over a range of 2\,$\mu$m, reaching a localization precision of about 10-20\,nm \cite{pavani2009three,zhou2018computational}.

By examining a rigorous model, numerical simulations and comparison with controlled measurements, we have shown how the use of iSCAT microscopy not only provides a high SNR and temporal resolution \cite{taylor2019Mini}, but it also provides a uniquely powerful avenue for recording three-dimensional trajectories with nanometer spatial precision. Here, the aberrated scattered field manifests as an extended and asymmetric field distribution along the optical axis. This distorted confinement also extends the distance over which the Gouy phase evolves. Together, these effects produce an iPSF consisting of an ample amount of light, allowing an unprecedented 3D tracking of a nanoparticle over the range of up to 10\,$\mu$m with nanometric precision. A similar result was also recently reported by formulating the effect of the lens in imaging an index-matched sample using an in-line digital holographic microscope \cite{leahy2020large}. These by far outperforms conventional dark-field and PSF-engineered fluorescence microscopies. 

The iPSF model presented here also opens new strategies for removal of the unwanted imaging background which accompanies the signal of interest. Previous works have exploited features unique to the PSF to separate the weak signal of nanoparticles from the speckle background of glass coverslips \cite{trueb2016robust, cheng2017background, aygun2019label} and for the dynamic backgrounds such as those encountered in living biological systems such as cells \cite{taylor2019interferometric}. Deconvolution with a deterministic iPSF or spectroscopic discrimination of the scattered field through a wavelength-dependent polarizability \cite{jacobsen2006interferometric} will further improve the separation of probe and background. Furthermore, while the axial extent of the iPSF is already large, we envisage improvements based on PSF engineering \cite{zhou2020precise}, permitting deep tissue imaging with interferometric sensitivity and precision. We speculate that additional aberrations originating from a highly scattering media, such as biological tissue, might also encode a wealth of information about the scatterer and its environment which could be retrieved over an extended axial range necessary for deep tissue imaging.

The model framework presented here can be extended to describe the position and orientation of non-symmetric particles such as nano-ellipsoids. It can also model the iPSF for a nano-scatterer illuminated by non-planar wavefronts and is, therefore, readily applicable to other common microscopy modalities such as confocal and bright-field transmission schemes. Furthermore, phase plates used to boost the detection contrast of particularly weak scatterers can be treated \cite{cole2017label,liebel2017ultrasensitive}. 

Another promising avenue for advanced image analysis aided by the iPSF model is the extraction of important physical quantities through machine learning. To benefit from fully automatic and robust deep learning techniques, one needs reliable physical models to generate ground truth iPSFs to train such networks. Our work provides a model which can generate data for use in training of supervised machine learning methods or in validating the outcomes of unsupervised machine learning. 

In conclusion, we have used experimental observation and numerical simulation to confirm that the vectorial diffraction model developed in this article accurately describes the operation of the popular wide-field reflection iSCAT microscope. We have shown that the interferometric point spread function, assisted by geometry-induced aberrations in the scattering wavefront, contains a wealth of information about the scattering source, allowing nanometrically accurate localization over an extended axial range in all three dimensions. We emphasize, however, that the radiation pattern of a scatterer is modified when situated at a sub-wavelength distance from the sample-substrate interface \cite{Lukosz:79}. This effect, which also modifies the iPSF profile, has been neglected in our current work, but it will be the topic of a sequel publication. 

\section*{Funding}
This project was funded by an Alexander von Humboldt professorship and postdoctoral fellowship, Research and Training Grant 1962 (Dynamic Interactions at Biological Membranes) of the German Research Foundation as well as by continuous support from the Max Planck Society.

\section*{Acknowledgments}
The authors would like to acknowledge the support of Alexander Gumann and Jan Renger for the nano-fabrication of coverslips, Goran Ahmed for fabrication of the quartz capillary tip, Korenobu Matsuzaki, Hsuan-Wei Liu and David Albrecht for help with experiments involving quartz fiber tips, as well as Mario Agio, Burak G\"urlek and Jahangir Nobakht for fruitful discussions. We are also grateful to Anna Kashkanova, Mahdi Mazaheri and Kiarash Kasaian for a careful reading of the manuscript and insightful comments.

\bibliography{references}

\begin{thebibliography}{10}
\newcommand{\enquote}[1]{``#1''}

\bibitem{lindfors2004detection}
K.~Lindfors, T.~Kalkbrenner, P.~Stoller, and V.~Sandoghdar, \enquote{Detection
  and spectroscopy of gold nanoparticles using supercontinuum white light
  confocal microscopy,} {\protect\JournalTitle{Physical Review Letters}}
  \textbf{93}, 037401 (2004).

\bibitem{taylor2019Mini}
R.~W. Taylor and V.~Sandoghdar, \enquote{Interferometric scattering microscopy:
  seeing single nanoparticles and molecules via rayleigh scattering,}
  {\protect\JournalTitle{Nano Letters}}  (2019).

\bibitem{taylor2019label}
R.~W. Taylor and V.~Sandoghdar, \enquote{Interferometric scattering (iscat)
  microscopy and related techniques,} in \emph{Label-Free Super-Resolution
  Microscopy,}  V.~Astratov, ed. (Springer International Publishing, Cham,
  2019), pp. 25--65.

\bibitem{young2019interferometric}
G.~Young and P.~Kukura, \enquote{Interferometric scattering microscopy,}
  {\protect\JournalTitle{Annual Review of Physical Chemistry}} \textbf{70},
  301--322 (2019).

\bibitem{spindler2016visualization}
S.~Spindler, J.~Ehrig, K.~K{\"o}nig, T.~Nowak, M.~Piliarik, H.~E. Stein, R.~W.
  Taylor, E.~Garanger, S.~Lecommandoux, I.~D. Alves \emph{et~al.},
  \enquote{Visualization of lipids and proteins at high spatial and temporal
  resolution via interferometric scattering (iscat) microscopy,}
  {\protect\JournalTitle{Journal of Physics D: Applied Physics}} \textbf{49},
  274002 (2016).

\bibitem{celebrano2011single}
M.~Celebrano, P.~Kukura, A.~Renn, and V.~Sandoghdar, \enquote{Single-molecule
  imaging by optical absorption,} {\protect\JournalTitle{Nature Photonics}}
  \textbf{5}, 95 (2011).

\bibitem{kukura2010single}
P.~Kukura, M.~Celebrano, A.~Renn, and V.~Sandoghdar, \enquote{Single-molecule
  sensitivity in optical absorption at room temperature,}
  {\protect\JournalTitle{The Journal of Physical Chemistry Letters}}
  \textbf{1}, 3323--3327 (2010).

\bibitem{piliarik2014direct}
M.~Piliarik and V.~Sandoghdar, \enquote{Direct optical sensing of single
  unlabelled proteins and super-resolution imaging of their binding sites,}
  {\protect\JournalTitle{Nature Communications}} \textbf{5}, 4495 (2014).

\bibitem{krishnan2010geometry}
M.~Krishnan, N.~Mojarad, P.~Kukura, and V.~Sandoghdar,
  \enquote{Geometry-induced electrostatic trapping of nanometric objects in a
  fluid,} {\protect\JournalTitle{Nature}} \textbf{467}, 692 (2010).

\bibitem{fringes2016situ}
S.~Fringes, M.~Skaug, and A.~W. Knoll, \enquote{In situ contrast calibration to
  determine the height of individual diffusing nanoparticles in a tunable
  confinement,} {\protect\JournalTitle{Journal of Applied Physics}}
  \textbf{119}, 024303 (2016).

\bibitem{huang2017virus}
Y.-F. Huang, G.-Y. Zhuo, C.-Y. Chou, C.-H. Lin, W.~Chang, and C.-L. Hsieh,
  \enquote{Coherent brightfield microscopy provides the spatiotemporal
  resolution to study early stage viral infection in live cells,}
  {\protect\JournalTitle{ACS Nano}} \textbf{11}, 2575--2585 (2017).

\bibitem{de2018revealing}
G.~de~Wit, D.~Albrecht, H.~Ewers, and P.~Kukura, \enquote{Revealing
  compartmentalized diffusion in living cells with interferometric scattering
  microscopy,} {\protect\JournalTitle{Biophysical Journal}} \textbf{114},
  2945--2950 (2018).

\bibitem{taylor2019interferometric}
R.~W. Taylor, R.~G. Mahmoodabadi, V.~Rauschenberger, A.~Giessl, A.~Schambony,
  and V.~Sandoghdar, \enquote{Interferometric scattering microscopy reveals
  microsecond nanoscopic protein motion on a live cell membrane,}
  {\protect\JournalTitle{Nature Photonics}} \textbf{13}, 480--487 (2019).

\bibitem{Lee07}
S.-H. Lee, Y.~Roichman, G.-R. Yi, S.-H. Kim, S.-M. Yang, A.~van Blaaderen,
  P.~van Oostrum, and D.~G. Grier, \enquote{Characterizing and tracking single
  colloidal particles with video holographic microscopy,}
  {\protect\JournalTitle{Optics Express}} \textbf{15}, 18275--18282 (2007).

\bibitem{cheong2010strategies}
F.~C. Cheong, B.~J. Krishnatreya, and D.~G. Grier, \enquote{Strategies for
  three-dimensional particle tracking with holographic video microscopy,}
  {\protect\JournalTitle{Optics Express}} \textbf{18}, 13563--13573 (2010).

\bibitem{gao2017optical}
D.~Gao, W.~Ding, M.~Nieto-Vesperinas, X.~Ding, M.~Rahman, T.~Zhang, C.~Lim, and
  C.-W. Qiu, \enquote{Optical manipulation from the microscale to the
  nanoscale: fundamentals, advances and prospects,}
  {\protect\JournalTitle{Light: Science \& Applications}} \textbf{6}, e17039
  (2017).

\bibitem{BradacTweezer}
C.~Bradac, \enquote{Nanoscale optical trapping: A review,}
  {\protect\JournalTitle{Advanced Optical Materials}} \textbf{6}, 1800005
  (2018).

\bibitem{de2012recent}
I.~De~Vlaminck and C.~Dekker, \enquote{Recent advances in magnetic tweezers,}
  {\protect\JournalTitle{Annual Review of Biophysics}} \textbf{41}, 453--472
  (2012).

\bibitem{leahy2020large}
B.~Leahy, R.~Alexander, C.~Martin, S.~Barkley, and V.~N. Manoharan,
  \enquote{Large depth-of-field tracking of colloidal spheres in holographic
  microscopy by modeling the objective lens,} {\protect\JournalTitle{Optics
  Express}} \textbf{28}, 1061--1075 (2020).

\bibitem{jacobsenbook}
V.~Jacobsen, E.~Klotzsch, and V.~Sandoghdar, \enquote{Interferometric detection
  and tracking of nanoparticles,} in \emph{Nano Biophotonics: Science and
  Technology,}  vol.~3 H.~Masuhara, S.~Kawata, and F.~Tokunaga, eds. (Elsevier,
  Amsterdam, 2007), chap.~9, pp. 143--159.

\bibitem{hwang2007interferometry}
J.~Hwang and W.~Moerner, \enquote{Interferometry of a single nanoparticle using
  the gouy phase of a focused laser beam,} {\protect\JournalTitle{Optics
  Communications}} \textbf{280}, 487--491 (2007).

\bibitem{feng2001physical}
S.~Feng and H.~G. Winful, \enquote{Physical origin of the gouy phase shift,}
  {\protect\JournalTitle{Optics Letters}} \textbf{26}, 485--487 (2001).

\bibitem{torok1997electromagnetic}
P.~T{\"o}r{\"o}k and P.~Varga, \enquote{Electromagnetic diffraction of light
  focused through a stratified medium,} {\protect\JournalTitle{Applied Optics}}
  \textbf{36}, 2305--2312 (1997).

\bibitem{haeberle2003focusing}
O.~Haeberl{\'e}, \enquote{Focusing of light through a stratified medium: a
  practical approach for computing microscope point spread functions. part i:
  Conventional microscopy,} {\protect\JournalTitle{Optics Communications}}
  \textbf{216}, 55--63 (2003).

\bibitem{bohren2008absorption}
C.~F. Bohren and D.~R. Huffman, \emph{Absorption and scattering of light by
  small particles} (John Wiley \& Sons, 2008).

\bibitem{aguet2009super}
F.~Aguet, S.~Geissb{\"u}hler, I.~M{\"a}rki, T.~Lasser, and M.~Unser,
  \enquote{Super-resolution orientation estimation and localization of
  fluorescent dipoles using 3-d steerable filters,}
  {\protect\JournalTitle{Optics Express}} \textbf{17}, 6829--6848 (2009).

\bibitem{Gibson1992b}
S.~F. Gibson and F.~Lanni, \enquote{{Experimental test of an analytical model
  of aberration in an oil-immersion objective lens used in three-dimensional
  light microscopy},} {\protect\JournalTitle{Journal of the Optical Society of
  America A}} \textbf{9}, 154 (1992).

\bibitem{richards1959electromagnetic}
B.~Richards and E.~Wolf, \enquote{Electromagnetic diffraction in optical
  systems, ii. structure of the image field in an aplanatic system,}
  {\protect\JournalTitle{Proc. R. Soc. Lond. A}} \textbf{253}, 358--379 (1959).

\bibitem{wolf1959electromagnetic}
E.~Wolf, \enquote{Electromagnetic diffraction in optical systems-i. an integral
  representation of the image field,} {\protect\JournalTitle{Proceedings of the
  Royal Society of London. Series A. Mathematical and Physical Sciences}}
  \textbf{253}, 349--357 (1959).

\bibitem{egner1999equivalence}
A.~Egner and S.~W. Hell, \enquote{Equivalence of the huygens--fresnel and debye
  approach for the calculation of high aperture point-spread functions in the
  presence of refractive index mismatch,} {\protect\JournalTitle{Journal of
  Microscopy}} \textbf{193}, 244--249 (1999).

\bibitem{pavani2009three}
S.~R.~P. Pavani, M.~A. Thompson, J.~S. Biteen, S.~J. Lord, N.~Liu, R.~J. Twieg,
  R.~Piestun, and W.~Moerner, \enquote{Three-dimensional, single-molecule
  fluorescence imaging beyond the diffraction limit by using a double-helix
  point spread function,} {\protect\JournalTitle{Proceedings of the National
  Academy of Sciences}} \textbf{106}, 2995--2999 (2009).

\bibitem{shechtman2014optimal}
Y.~Shechtman, S.~J. Sahl, A.~S. Backer, and W.~Moerner, \enquote{Optimal point
  spread function design for 3d imaging,} {\protect\JournalTitle{Physical
  Review Letters}} \textbf{113}, 133902 (2014).

\bibitem{stein2017production}
H.~Stein, S.~Spindler, N.~Bonakdar, C.~Wang, and V.~Sandoghdar,
  \enquote{Production of isolated giant unilamellar vesicles under high salt
  concentrations,} {\protect\JournalTitle{Frontiers in Physiology}} \textbf{8},
  63 (2017).

\bibitem{Kaller2020GNP}
M.~Kaller, A.~Kashkanova, and \textit{et al.}, }}  (in preparation).

\bibitem{parthasarathy2012rapid}
R.~Parthasarathy, \enquote{Rapid, accurate particle tracking by calculation of
  radial symmetry centers,} {\protect\JournalTitle{Nature Methods}} \textbf{9},
  724 (2012).

\bibitem{Lin2020}
{Lin, Shupei and He, Yong and Robert, Hadrien Marc Louis and Li, Hong and
  Zhang, Pu and Piliarik, Marek and Chen, Xue-Wen}, \enquote{{Multiscale
  Modeling and Analysis for High-fidelity Interferometric Scattering
  Microscopy},} \url{https://arxiv.org/abs/2004.10575v3}.

\bibitem{deschout2014precisely}
H.~Deschout, F.~C. Zanacchi, M.~Mlodzianoski, A.~Diaspro, J.~Bewersdorf, S.~T.
  Hess, and K.~Braeckmans, \enquote{Precisely and accurately localizing single
  emitters in fluorescence microscopy,} {\protect\JournalTitle{Nature Methods}}
  \textbf{11}, 253 (2014).

\bibitem{zhou2019advances}
Y.~Zhou, M.~Handley, G.~Carles, and A.~R. Harvey, \enquote{Advances in 3d
  single particle localization microscopy,} {\protect\JournalTitle{APL
  Photonics}} \textbf{4}, 060901 (2019).

\bibitem{hannel2018machine}
M.~D. Hannel, A.~Abdulali, M.~O’Brien, and D.~G. Grier,
  \enquote{Machine-learning techniques for fast and accurate feature
  localization in holograms of colloidal particles,}
  {\protect\JournalTitle{Optics Express}} \textbf{26}, 15221--15231 (2018).

\bibitem{shao2020machine}
S.~Shao, K.~Mallery, S.~S. Kumar, and J.~Hong, \enquote{Machine learning
  holography for 3d particle field imaging,} {\protect\JournalTitle{Optics
  Express}} \textbf{28}, 2987--2999 (2020).

\bibitem{lloyd1982least}
S.~Lloyd, \enquote{Least squares quantization in pcm,}
  {\protect\JournalTitle{IEEE transactions on information theory}} \textbf{28},
  129--137 (1982).

\bibitem{david2007vassilvitskii}
D.~Arthur and S.~Vassilvitskii, \enquote{K-means++: The advantages of careful
  seeding,} in \emph{18th annual ACM-SIAM symposium on Discrete algorithms
  (SODA), New Orleans, Louisiana,}  (2007), pp. 1027--1035.

\bibitem{rousseeuw1987silhouettes}
P.~J. Rousseeuw, \enquote{Silhouettes: a graphical aid to the interpretation
  and validation of cluster analysis,} {\protect\JournalTitle{Journal of
  Computational and Applied Mathematics}} \textbf{20}, 53--65 (1987).

\bibitem{weisenburger2015light}
S.~Weisenburger and V.~Sandoghdar, \enquote{Light microscopy: an ongoing
  contemporary revolution,} {\protect\JournalTitle{Contemporary Physics}}
  \textbf{56}, 123--143 (2015).

\bibitem{zhou2018computational}
Y.~Zhou, P.~Zammit, G.~Carles, and A.~R. Harvey, \enquote{Computational
  localization microscopy with extended axial range,}
  {\protect\JournalTitle{Optics Express}} \textbf{26}, 7563--7577 (2018).

\bibitem{trueb2016robust}
J.~T. Trueb, O.~Avci, D.~Sevenler, J.~H. Connor, and M.~S. {\"U}nl{\"u},
  \enquote{Robust visualization and discrimination of nanoparticles by
  interferometric imaging,} {\protect\JournalTitle{IEEE Journal of Selected
  Topics in Quantum Electronics}} \textbf{23}, 394--403 (2016).

\bibitem{cheng2017background}
C.-Y. Cheng and C.-L. Hsieh, \enquote{Background estimation and correction for
  high-precision localization microscopy,} {\protect\JournalTitle{ACS
  Photonics}} \textbf{4}, 1730--1739 (2017).

\bibitem{aygun2019label}
U.~Aygun, H.~Urey, and A.~Y. Ozkumur, \enquote{Label-free detection of
  nanoparticles using depth scanning correlation interferometric microscopy,}
  {\protect\JournalTitle{Scientific Reports}} \textbf{9}, 1--8 (2019).

\bibitem{jacobsen2006interferometric}
V.~Jacobsen, P.~Stoller, C.~Brunner, V.~Vogel, and V.~Sandoghdar,
  \enquote{Interferometric optical detection and tracking of very small gold
  nanoparticles at a water-glass interface,} {\protect\JournalTitle{Optics
  Express}} \textbf{14}, 405--414 (2006).

\bibitem{zhou2020precise}
Y.~Zhou and G.~Carles, \enquote{Precise 3d particle localization over large
  axial ranges using secondary astigmatism,} {\protect\JournalTitle{Optics
  Letters}} \textbf{45}, 2466--2469 (2020).

\bibitem{cole2017label}
D.~Cole, G.~Young, A.~Weigel, A.~Sebesta, and P.~Kukura, \enquote{Label-free
  single-molecule imaging with numerical-aperture-shaped interferometric
  scattering microscopy,} {\protect\JournalTitle{ACS Photonics}} \textbf{4},
  211--216 (2017).

\bibitem{liebel2017ultrasensitive}
M.~Liebel, J.~T. Hugall, and N.~F. van Hulst, \enquote{Ultrasensitive
  label-free nanosensing and high-speed tracking of single proteins,}
  {\protect\JournalTitle{Nano Letters}} \textbf{17}, 1277--1281 (2017).

\bibitem{Lukosz:79}
W.~Lukosz, \enquote{Light emission by magnetic and electric dipoles close to a
  plane dielectric interface. iii. radiation patterns of dipoles with arbitrary
  orientation,} {\protect\JournalTitle{J. Opt. Soc. Am.}} \textbf{69},
  1495--1503 (1979).

\end{thebibliography}
\end{document}